\documentclass[12pt,dvips]{article}

\usepackage{epsfig}
\newcommand{\be}{\begin{eqnarray}}
\newcommand{\ee}{\end{eqnarray}}
\newcommand{\ben}{\begin{eqnarray*}}
\newcommand{\een}{\end{eqnarray*}}
\begin{document}

\def\J{$J/\psi$}
\def\j{J/\psi}
\def\P{$\psi'$}
\def\p{\psi'}
\def\U{$\Upsilon$}
\def\u{\Upsilon}
\def\C{c{\bar c}}
\def\cg{c{\bar c}\!-\!g}
\def\bg{b{\bar b}\!-\!g}
\def\b{b{\bar b}}
\def\q{q{\bar q}}
\def\Q{Q{\bar Q}}
\def\F{$\Phi$}
\def\f{\Phi}
\def\T{$\hat t$}
\def\CG{$c{\bar c}\!-\!g$~}
\def\d{$d{\hat{\sigma}}$}
\def\t{\tau}
\def\E{\epsilon}
\def\a{\alpha_s}

\def\T{$T_f$}
\def\n{$n_b$}

\def\L{\cal L}

\def\e{\epsilon}

\def\lsim{\raise0.3ex\hbox{$<$\kern-0.75em\raise-1.1ex\hbox{$\sim$}}}
\def\gsim{\raise0.3ex\hbox{$>$\kern-0.75em\raise-1.1ex\hbox{$\sim$}}}

%
\def\CMP{{ Comm.\ Math.\ Phys.\ }}
\def\NP{{ Nucl.\ Phys.\ }}
\def\PL{{ Phys.\ Lett.\ }}
\def\PR{{ Phys.\ Rev.\ }}
\def\PRep{{ Phys.\ Rep.\ }}
\def\PRL{{ Phys.\ Rev.\ Lett.\ }}
\def\RMP{{ Rev.\ Mod.\ Phys.\ }}
\def\ZP{{ Z.\ Phys.\ }}

\def\etal{{\sl et al.}}

\thispagestyle{empty}
\begin{flushright}

CERN-TH/96-328\\
BI-TP 96/53
\end{flushright}
\vskip1.6cm

\centerline{\large{\bf A Quantitative Analysis of}}
\vskip0.5cm
\centerline{\large{\bf Charmonium Suppression in Nuclear Collisions}}
\vskip1.2cm
\centerline{\bf D. Kharzeev$^{1}$, C. Louren\c{c}o$^{2}$, M.
Nardi$^{1,3}$ and H. Satz$^{1,4}$}
\vskip0.5cm
\centerline{{\it 1 Fakult\"at f\"ur Physik, Universit\"at Bielefeld,
D-33501 Bielefeld,
 Germany}}
\centerline{{\it 2 PPE Division, CERN, CH-1211 Geneva 23, Switzerland}}
\centerline{{\it 3 INFN Torino, Universit\`a di Torino, I-10125 Torino,
Italy}}
\centerline{{\it 4 Theory Division, CERN, CH-1211 Geneva 23,
Switzerland}}
\vskip1.5cm
\begin{abstract}

~~~Data from \J~and \P~production in $p-A$ collisions are used to
determine
the cross section for absorption of pre-resonance charmonium in nuclear
matter. The \J~suppression in $O-Cu$, $O-U$ and $S-U$ collisions is
fully reproduced by the corresponding nuclear absorption, while $Pb-Pb$
collisions show an additional suppression increasing with centrality.
We study the onset of this change in terms of hadronic comover
interactions and conclude that so far no conventional hadronic
description can consistently account for all data. Deconfinement,
starting at a critical point determined by central $S-U$ collisions,
is in accord with the observed suppression pattern.
\end{abstract}

\vskip1.2cm
\begin{flushleft}
CERN-TH/96-328\\
BI-TP 96/53\\
November 1996
\end{flushleft}
\newpage

\noindent
{\bf 1.\ Introduction}
\vskip0.5cm
The suppression of \J~production in nucleus-nucleus collisions was
proposed ten years ago as a signal of colour deconfinement \cite{Matsui}.
When a suppression was observed in $O-U$ and later in $S-U$
collisions \cite{NA38}, it was noted that already $p-A$ interactions
result in reduced \J~production \cite{Capella}. Following recent studies
of charmonium production in hadron-hadron interactions
\cite{Bodwin,Braaten}, \J~suppression in $p-A$ collisions can be
understood as absorption of a pre-resonance $\C-g$ state in nuclear
matter \cite{KS6}. The onset of colour deconfinement must then lead to
an additional suppression of \J~production, beyond the mentioned
nuclear absorption; in other words, it must result in a reduction
beyond what is already observed in $p-A$ interactions \cite{Gerschel}.
\par
The ``anomalous" \J~suppression recently reported by the NA50
collaboration \cite{Paula}-\cite{Carlos} therefore triggered considerable
excitement and already quite a number of tentative explanations
\cite{Blaizot}-\cite{Cassing}. While we fully share this excitement, we
believe that present data are sufficient to carry out a {\sl
quantitative} analysis of charmonium suppression in nuclear
collisions. There now exist quite extensive results on \J~production
in $p-A$ interactions for the determination of nuclear
absorption, and the introduction of a zero degree calorimeter in the
experimental set-up of NA50 specifies the underlying collision
geometry much better than before. Combining the two in a systematic
analysis leads to some clear-cut conclusions and also shows what further
data is necessary to corroborate these.
\par
We begin our analysis, in Section 2, with a study of pre-resonance
charmonium absorption in $p-A$ collisions, using the Glauber theory of
nuclear interactions. Apart from the well-established nuclear profile,
this formalism contains as only parameter the absorption cross section
$\sigma_{abs}$ in nuclear matter, which can thus be determined with a
precision limited only by that of the $p-A$ data.
Turning to the collision geometry of nucleus-nucleus interactions, we
establish in Section 3 the relation between the centrality of the
collision, the associated transverse energy and the associated number
of spectator nucleons.
\par
Using these results, we show in Section 4 that
pre-resonance absorption fully accounts for the \J~suppression
observed in $O-Cu$, $O-U$ and $S-U$ collisions. In contrast, $Pb-Pb$
collisions show considerable \J~suppression beyond nuclear absorption.
For the \P, there is increased absorption already in $S-U$ collisions.
\par
In Section 5, we then investigate if
the observed \J~and \P~ suppression can be understood in
``conventional" terms, i.e., as dissociation in a system of hadronic
comovers produced in the course of the collision. We find that the
anomalous \J~suppression in $Pb-Pb$ and only nuclear absorption in
$S-U$ collisions make this impossible.
In Section 6, we consider charmonium suppression by
local colour deconfinement, assumed to begin at a critical density
just above that of central $S-U$ collisions. The resulting
suppression pattern agrees with the $Pb-Pb$ data.
\vskip0.5cm
\noindent
{\bf 2.\ Nuclear Absorption in $p-A$ Collisions}
\vskip0.5cm
Charmonium production in hadron-hadron collisions proceeds through
parton fusion, at high energies predominantly gluon fusion, to form a
$\C$ pair \cite{HP}. Because of the high mass $m_c$ of the charm quark,
this process occurs almost instantaneously, with a formation time $\t_c
\simeq (2m_c)^{-1} \simeq 0.07$ fm in the $\C$ rest frame. The $\C$ pair
is generally in a coloured state; to neutralize its colour and form
a singlet $\C$ state \J~or \P~requires a considerably longer
time; virtuality estimates \cite{KS6} give about 0.3 fm, again in the
rest frame of the $\C$. The analysis of recent experiments
\cite{Fermilab} has shown the role of higher Fock space components in
charmonium hadroproduction \cite{Bodwin,Braaten}. For production
at low transverse momentum this suggests that the pre-resonance state
of charmonia in the first 0.3 fm is a colour octet $\C$ coupled with
a soft collinear gluon to neutralise the overall colour of this $\C-g$
system \cite{KS6}. In the presently accessible kinematic
region of \J~production by $p-A$ collisions ($x_F\geq 0$), the target
nucleus sees only the passage of the pre-resonance state; physical
charmonium states are formed outside the nucleus.
\par
The suppression of \J~production in $p-A$ collisions should thus be
understood as pre-resonance absorption in normal nuclear matter. The
size of the pre-resonance state is determined by the charm
quark mass and the confinement scale; hence it is the same for \J~and
\P. This accounts naturally for the equal suppression observed for the
two states, which would be impossible for physical resonances whose
transverse areas differ by more than a factor three.
Theoretical estimates suggest for the absorption
cross section of the $\C-g$ on nucleons $\sigma_{abs} \simeq 6 - 7$ mb
\cite{KS6}. We want to see to what extent it is possible to determine
this cross section in a systematic study of presently available $p-A$
data.
\par
In Glauber theory, the survival probability for a \J~produced in a
$p-A$ collision is given by
\ben
S^{pA}_{Gl} = \sigma(pA\to \psi)/ A \sigma(pN \to \psi) =
\nonumber
\een
\be
= \int d^2b~ dz~\rho_A(b,z) \exp\left\{-(A-1)\int_z^{\infty} dz'
\rho_A(b,z')~\sigma_{abs} \right\}. \label{1}
\ee
Here $\rho_A$ is the nuclear density distribution, for which we take the
standard three-parameter Woods-Saxon form with parameters as tabulated
\cite{deJager}; it is normalized to unity, with
\be
\int d^2b~  dz~ \rho_A(b,z) = 1. \label{2}
\ee
In Eq.\ (\ref{1}), the integration runs over the impact parameter $b$
of the incident proton and the production point $z$ of the pre-resonance
state; the integral in the weight function covers the path of this
state. The suppression is thus fully determined by the absorption
cross section $\sigma_{abs}$ in nuclear matter.
\par
Experimentally, the \J~survival probability
\be
S^{exp}_{pA} \equiv \sigma^{exp}_{pA}~/~ A \sigma^{exp}_{pN},
\label{3}
\ee
is determined by comparing $p-A$ with $p-p$ or $p-D$ results. We start
with the latter, since here there are data at 200 [NA38], 450 [NA38/51]
and 800 GeV/c [E772]; for a recent compilation, see \cite{Carlos}. In
Table 1a, we list the corresponding survival probabities. The errors of
the 800 GeV/c data \cite{E772} are significantly smaller than
those at the other energies because here $p-A$ and $p-D$ collisions
were studied in the same experiment. Using the 450 GeV/c NA51 $p-D$
cross section \cite{NA51} to normalize the 450 GeV/c NA38 $p-A$ data
\cite{Fredj} results in larger errors. To normalize the 200 GeV/c
$p-A$ data, the 450 GeV/c $p-D$ cross section moreover has to be
rescaled \cite{Carlos}, using the experimentally known energy
dependence of \J~production. -- We now use the A-dependence of these
results to determine the nuclear absorption cross section for
pre-resonance charmonium.
\par
For the combined data in Table 1a we obtain the best fit with
\be
\sigma_{abs} = 7.3 \pm 0.6~ {\rm mb}; \label{4}
\ee
it gives a $\chi^2/d.f.\simeq 1.4$, with the error corresponding to a
confidence level of 95\% (see Fig.\ 1). Analyzing the 800, 450 and
200 GeV/c data separately gives clear $\chi^2$ minima at 7.3, 7.4 and
7.1 mb, respectively. The results thus are fully compatible; they
indicate that the $A$-dependence does not vary with incident energy in
the presently studied range. This will presumably change when the beam
energies become much higher (RHIC and LHC), since then nuclear shadowing
of the gluon distributions is expected to enhance the suppression \cite{GS2}. The
survival probabilities obtained with the cross section (\ref{4}) are
included in Table 1a. In Fig.\ 2, we compare the experimental and the
nuclear absorption results; the agreement is very good.
\par
We have also carried out the corresponding analysis of the 450 and
200 GeV/c $p-A$ results with the NA51 $p-p$ data as reference.
The resulting $\sigma_{abs} = 6.4 \pm 0.6$ mb is somewhat lower, but
still compatible with Eq.\ (\ref{4}). However, here the large errors of
the data make it difficult to define limits, so that the quoted range
in this case corresponds to twice the minimum $\chi^2$.

\begin{table}
\begin{center}
\begin{tabular}{|c||c||c||c|}
\hline       &   &   & \\
& $P_{beam}$& $S_{pA}^{exp}$ & $S_{Gl}$\\
& [GeV/c]&  & (7.3 $\pm$0.6 mb)\\
&    & &  \\
\hline\hline
p-C  & 800 & 0.851 $\pm$ 0.013 & 0.867 $\pm$ 0.010 \\ \hline
p-Ca & 800 & 0.806 $\pm$ 0.009 & 0.784 $\pm$ 0.014 \\ \hline
p-Fe & 800 & 0.756 $\pm$ 0.010 & 0.753 $\pm$ 0.015 \\ \hline
p-W  & 800 & 0.619 $\pm$ 0.013 & 0.650 $\pm$ 0.021 \\
\hline\hline
p-C  & 450 & 0.85 $\pm$ 0.10 & 0.87 $\pm$ 0.01 \\ \hline
p-Al & 450 & 0.76 $\pm$ 0.09 & 0.81 $\pm$ 0.02 \\ \hline
p-Cu & 450 & 0.75 $\pm$ 0.08 & 0.74 $\pm$ 0.02 \\ \hline
p-W  & 450 & 0.67 $\pm$ 0.08 & 0.65 $\pm$ 0.02 \\
\hline\hline
p-Cu & 200 &0.75 $\pm$ 0.17 & 0.74 $\pm$ 0.02 \\ \hline
p-W  & 200 &0.65 $\pm$ 0.05 & 0.65 $\pm$ 0.02 \\ \hline
p-U  & 200 &0.63 $\pm$ 0.13 & 0.61 $\pm$ 0.02 \\ \hline
\end{tabular}
\vskip 0.8cm
(a) $J/\psi$ suppression from $pA/pD$ [2,10,22]
\vskip 1.2cm
\begin{tabular}{|c||c||c||c|}
\hline       &   &   & \\
& $P_{beam}$& $S_{pA}^{exp} /S_{pC}^{exp}$ & $S_{Gl}^{pA}/S_{Gl}^{pC}$\\
& [GeV/c]&  & (7.3 $\pm$0.6 mb)\\
&    & &  \\
\hline\hline
Ca/C & 800 & 0.947 $\pm$ 0.024 & 0.904 $\pm$ 0.007 \\ \hline
Fe/C & 800 & 0.888 $\pm$ 0.025 & 0.869 $\pm$ 0.009 \\ \hline
W/C  & 800 & 0.727 $\pm$ 0.026 & 0.750 $\pm$ 0.015 \\
\hline\hline
Al/C & 450 & 0.925 $\pm$ 0.055 & 0.931 $\pm$ 0.005 \\ \hline
Cu/C & 450 & 0.855 $\pm$ 0.035 & 0.859 $\pm$ 0.009 \\ \hline
W/C  & 450 & 0.761 $\pm$ 0.043 & 0.750 $\pm$ 0.015 \\
\hline
\end{tabular}
\vskip 0.8cm
(b) $J/\psi$ suppression ratios $pA/pC$ [22,23]
\end{center}
\vskip 0.5cm
\caption{$J/\psi$ suppression in $p-A$ collisions}
\end{table}

So far, the nuclear suppression cross section was obtained with $p-D$
or $p-p$ data as reference. There is an alternative method which does
not require these $p-N$ values. The nuclear target ratios $S_{pA}/
S_{pB}$ depend directly on $\sigma_{abs}$. We will here compare
the results on different targets to those on carbon, since for this
case there exist also 450 GeV/c data with relatively small errors
\cite{Fredj}. The ratios at 800 and 450 GeV/c are listed in Table 1b;
they provide the best fit for $\sigma_{abs} = 7.0 \pm 1.5$,
with $\chi^2/d.f. \simeq 0.9$ and the error determined by a 95\%
confidence level. The $\chi^2$ distribution is included in Fig.\ 1;
again the different experiments also have their individual $\chi^2$
minima at 7 mb. We thus find that the value of $\sigma_{abs}$ obtained
through nuclear ratios is in good agreement with that obtained from
Table 1a. Since the value (\ref{4}) is the most precise, we will use it
in the remainder of our analysis.
\par
Next we turn to a comparison of \J~and \P~production in $p-A$
collisions. If the suppression observed in these reactions is indeed due
to pre-resonance absorption, the ratio \P/(\J) should remain for all $A$
at its value in nucleon-nucleon interactions, where it was found to be
$(1.81 \pm 0.38)\times 10^{-2}$ and independent of the c.m.s. collision
energy from 20 to almost 2000 GeV \cite{HP}. Fitting the $p-A$ results
compiled in \cite{Carlos} by the form
\be
{B(\psi'\to \mu^+\mu^-)~ \sigma_{pA \to \psi'} \over B(\j \to
\mu^+\mu^-)~\sigma_{pA \to \j}}
= c~A^{(\alpha_{\psi'}-\alpha_{\j})},
\label{5}
\ee
we obtain $(\alpha_{\psi'}-\alpha_{\psi}) = 0.00 \pm 0.02$ with a 95\%
confidence level; this rules out variations of more than 10\% between
$p-p$ and $p-U$ collisions. The value $c=(1.74 \pm 0.07)\times 10^{-2}$
agrees very well with the mentioned nucleon-nucleon average.
\par
We thus conclude that the attenuation of \J~and \P~in $p-A$ collisions
is quantitatively well described as the absorption of a pre-resonance
charmonium state in nuclear matter, with $\sigma_{abs}= 7.3 \pm 0.6$ mb.
The $A$-independence of \P/(\J) establishes that the absorption occurs
before the formation of physical resonances.
\vskip0.5cm
\noindent
{\bf 3.\ The Centrality Dependence of Nuclear Collisions}
\vskip0.5cm
In nucleus-nucleus collisions, charmonium production can be measured
as function of the centrality of the collision, and hence we have to
calculate the \J~survival probability at fixed impact parameter $b$. For
an $A-B$ collision, it is given by

\ben
{ d S_{Gl}(b) \over d^2 b} =
{1 \over A B~ \sigma(NN \to \psi)}\left[{d \sigma(AB
\to \psi) \over d^2 b} \right] =  \nonumber
\een
\ben
=  \int d^2s~ dz~ dz'~\rho_A(\vec{s},z)~ \rho_B(\vec{b}-\vec{s},z')~
 \exp\left\{-(A-1)\int_z^{\infty} dz_A~ \rho_A(\vec{s},z_A)~
 \sigma_{abs} \right\}& \nonumber
\een
\be
~~~~\times \exp\left\{-(B-1)\int_{z'}^{\infty} dz_B~
\rho_B(\vec{b}-\vec{s},z_B)~
\sigma_{abs} \right\}. \label{6}
\ee
Here $\vec{s}$ specifies the position of the production point in a plane
orthogonal to the collision axis, while $z$ and $z'$ give the position
of this point within nucleus $A$ and within nucleus $B$, respectively.
The nuclear density distributions $\rho_A$ and $\rho_B$ are defined as
above. Note that $d S_{Gl}/d^2b$ is normalized such that when
$\sigma_{abs}=0$,
\be
\int d^2b~\left\{{d S_{Gl}(b; \sigma_{abs}=0)\over d^2b}\right\} = 1.
\label{7}
\ee
To obtain a normalized probability at fixed impact parameter $b$,
we have to divide $[d S_{Gl}(b)/d^2b]$ by
$[d S_{Gl}(b;\sigma_{abs}=0)/d^2b]$.
\par
Experimentally, the centrality of the collision is determined by a
calorimetric measurement of the associated transverse energy $E_T$, i.e.,
the total energy emitted in the form of hadrons in a plane orthogonal to
the beam axis. The more central a collision is, the more nucleons from
target and projectile will participate in the interaction, and hence
the more secondary hadrons will be produced. Starting with the $Pb$-beam
experiments at CERN, the NA50 collaboration uses in addition a zero
degree calorimeter; it counts the number of projectile nucleons which
have not participated in the collision and therefore provides a further
check on the centrality of the reaction. To compare our calculations
with the measured survival probability $S_{AB}^{exp}(E_T)$, we thus have
to establish and test a correspondence between the impact parameter $b$,
the transverse energy $E_T$, and the energy $E_Z$ reaching the zero
degree calorimeter.
\par
This correlation is given in terms of the number of ``wounded" nucleons
\cite{Bialas}. Any given nucleon in a $p-A$ or $A-B$ collision is
designated as wounded when it has interacted inelastically one or
more times with one or more other nucleons, and fragmentation of
such wounded nucleons eventually
produces the observed secondary hadrons. At high energies, a projectile
nucleon in
general passes the entire target nucleus in a time much shorter than
needed to transfer information across its own spatial extension, and
vice versa for the target nucleons. For this reason, the number of times
a nucleon is wounded does not affect the subsequent production of
hadronic secondaries in its fragmentation. A given wounded nucleon
produces on the average $N_h$ hadrons, and if each of these hadrons
has an average transverse energy $q_h$, then each wounded nucleon will
contribute $q \equiv N_h q_h$ to the overall transverse energy
produced in the collision. We thus have the relation
\be
E_T(b)~ =~q~N_w(b) \label{8}
\ee
between the average number $N_w$ of nucleons wounded in a
collision at fixed impact parameter $b$ and the associated average
transverse energy $E_T$ produced in that collision. In general, $q$ will
depend on the incident energy in the same way as the average
multiplicity. Besides this, in the analysis of specific experimental
results,
$q$ depends on the details of the detector, in particular on the
rapidity and transverse momentum range in which the produced
secondaries are measured. Note that relation (\ref{8}) is no
longer valid when $E_T > E_T(b=0)$, i.e., when $E_T$ is increased by
multiplicity fluctuations at $b=0$.
\par
The average number of wounded nucleons in an $A-B$ collision at impact
parameter $b$ is given by

\be
N_w^{AB}(b)~= & A \int d^2 s~ T_A(\vec{s}) \left\{ 1-[1-\sigma_N
T_B(\vec{s}-\vec{b})]^B\right\} + & \nonumber \\
+ & B \int d^2 s~ T_B(\vec{s}-\vec{b}) \left\{ 1-[1-\sigma_N
T_A(\vec{s})]^A\right\}. & \label{9}
\ee
Here $\sigma_N\simeq 30$ mb denotes the inelastic nucleon-nucleon
cross section without the diffractive production component, and
$T_A(\vec{s}) =\int dz~\rho_A(z,\vec{s})$ the nuclear profile
function; the $\vec{s}$-integration runs again over a plane orthogonal
to the collision axis. The two terms in Eq.\ (\ref{9}) thus correspond
to the wounding of nucleons from nucleus $A$ in its passage of nucleus
$B$ and vice versa.
\par
Integrating Eq.\ (\ref{9}) over impact parameter yields
\ben
\int d^2 b~ N_w^{AB}(b) =
\een
\be
A \int d^2 s~ \left\{ 1-[1-\sigma_N T_B(\vec{s})]^B\right\}
+ B \int d^2 s~ \left\{ 1-[1-\sigma_N T_A(\vec{s})]^A\right\},
\label{10}
\ee
since $\int d^2 b~ T(\vec{s}-\vec{b}) = 1$ for any $\vec{s}$, see
Eq.\ (\ref{2}). Since the inelastic cross section for nucleon--nucleus
collisions is given by
\be
\sigma_A
= \int d^2 s~ \left\{ 1-[1-\sigma_N T_A(\vec{s})]^A\right\}, \label{11}
\ee
Eq.\ (\ref{10}) can be re-written in the familiar form
\cite{Bialas}
\be
\int d^2 b~ N_w^{AB}(b) = (A \sigma_B + B \sigma_A). \label{12}
\ee
Because of fluctuations in the number of wounded nucleons and in the
transverse energy of the secondaries that each wounded nucleon produces,
there will be corresponding fluctuations in the relation between $E_T$
and $b$. We assume the dispersion $D$ in the number of wounded nucleons
to be proportional to $\sqrt{N}_w$,
\be
D^2_{AB}~ =~ a~ N_w^{AB}(b), \label{13}
\ee
with a dimensionless universal physical parameter $a$ found to be about
one in $A-B$ collisions \cite{Bialas}. The $E_T-b$ correlation function
(see Eq.\ (\ref{8})) can then be written as
\be
P_{AB}(E_T,b)~=~{1 \over \sqrt{2\pi q^2a  N^{AB}_w(b)}}
\exp\left\{-{[E_T - q N^{AB}_w(b)]^2 \over 2q^2 a
N^{AB}_w(b)}\right\};
\label{14}
\ee
with
\be
\int dE_T ~P_{AB}(E_T,b) = 1,\label{15}
\ee
so that it is normalized at fixed $b$.
\par
We now have to check whether the resulting form indeed accounts for the
observed $E_T$-distributions and fix the fluctuation dispersion $a$.
These general questions are best answered using the minimum bias data
of the NA35/49 collaboration \cite{Stock}; that will also allow us to
later address charmonium production data with as well-determined a
form as possible. The minimum bias cross section as function of $E_T$
is given by
\be
{d\sigma_{AB} \over dE_T} = \int d^2 b~[1-P_0(b)]~P_{AB}(E_T,b),
\label{16}
\ee
where $P_0(b)=[1-\sigma_N T_{AB}(b)]^{AB}$ is the probability for no
interaction.  Hence $[1-P_0(b)]$ cuts off the integral at large $b$,
when interactions become impossible (for a hard sphere model, this
would be at $b= R_A+R_B$).
In Fig.\ 3, we compare Eq.\ (\ref{16}) to the measured minimum bias
$E_T$ distributions from $S-Au$ and $Pb-Pb$ collisions \cite{Stock}.
Note that the $S-Au$ data were measured in a larger rapidity window
than the $Pb-Pb$ data; they were then reanalysed to obtain the $S-Au$
$E_T$-distribution in the $Pb-Pb$ rapidity window \cite{Spiros}.
Setting $a=1$ \cite{Bialas} and with $q=1.5$ for $S-Au$ and $q=1.4$
for $Pb-Pb$ interactions, Eq.\ (\ref{16}) is seen to provide an
excellent description of the measured $E_T$ distributions. Since the
acceptance is the same in both cases, the difference in the observed
$q$-values is expected to reflect the difference in collision
energies; it thus provides us with some estimate of the energy
dependence.
\par
The distribution of the transverse energy associated to charmonium
production will differ from that just discussed, since
the $A-B$ dependence of hard processes is determined by the number of
collisions, rather than by the number of wounded nucleons. When a
nucleon, in passing a nucleus, gets wounded several times, 
at present energies this does not
change the multiplicity of secondaries in its subsequent hadronisation;
but each of the collisions can in principle produce a hard dilepton
pair or a heavy quark state. The number of collisions in an $A-B$
interaction at impact parameter
$b$ is given by
\be
N^{AB}_c(b)~=~A B~T_{AB}(b)~\sigma_N, \label{17}
\ee
where $T_{AB}(\vec{b})=\int d^2s~T_A(\vec{s})T_B(\vec{b}-\vec{s})$ is
the nuclear overlap function, normalized to unity: $\int d^2b~
T_{AB}(b) = 1$. Using it, we can define the ``conditional"
$E_T$-distribution, associated with a hard process such as Drell-Yan
or charmonium production,
\be
{d\sigma^h_{AB} \over dE_T} ={\sigma_N^h \over \sigma_N}
\int d^2 b~ N_c^{AB}(b)~P_{AB}(E_T,b). \label{18}
\ee
It is seen to be correctly normalized, giving
\be
\sigma_{AB}^h = {\sigma_N^h \over \sigma_N} \int d^2b~ N_c(b) =
AB~\sigma_N^h, \label{19}
\ee
where $\sigma^h_N$ is the cross section for the corresponding hard
process in an elementary $NN$ collision.
\par
Before we can compare our charmonium survival probabilities at fixed
$E_T$ with data, we thus have to check that
also the conditional $E_T$-distributions are correctly reproduced with
our $E_T-b$ assignment. Since the acceptance in the NA38/50 set-up
differs from the minimum bias experiments studied above, we have to
determine once again the mean energy $q$ measured per wounded nucleon;
the dispersion in $N_w$ is retained as determined above. We see in Fig.\
4 that with $q=0.75$ for $S-U$ and $q=0.40$ for $Pb-Pb$ collisions the
conditional $E_T$ distributions are reproduced very well. We
have included in our distribution the measured \J~suppression relative
to the Drell-Yan continuum, since the presently available distributions
are dominated by \J~rather than Drell-Yan events.
At low $E_T$, the data fall below the
calculated distribution; this is in both cases due to the $E_T$
acceptance profile of the experiment \cite{Carlos-Dis}. The larger
difference between the two $q$-values here, as compared to minimum bias,
is due to different rapidity coverages of the NA38 and NA50
calorimeters.
\par
As already mentioned, the $Pb-Pb$ experiment provides full collision
geometry determination by means of a zero degree calorimeter (ZDC),
which
measures at each $E_T$ the associated number of projectile spectators
-- those projectile nucleons which reach the ZDC with their full
initial energy $E_{in}=158$ GeV. This additional information uniquely
identifies the peripherality of the collisions leading to the measured
charmonium production; qualitatively, the NA50 results are quite similar
to the minimum bias results of WA80 \cite{Albrecht}. The measured
$E_T-E_Z$ correlation \cite{Scomparin} in Fig.\ 5 indicates, e.g., that for
transverse energy $E_T \simeq 90$ GeV approximately half the
projectile nucleons remain spectators; therefore
the impact parameter in this case must be around $b \sim R_{Pb}$.
In general, the number of projectile spectators is $A - N_w^A$,
so that $E_Z=(A-N_w^A)E_{in}$. Using $E_T=q~N_w=2q~N_w^A$, this leads
to the $E_T-E_Z$ correlation shown in Fig.\ 5 (labelled `Glauber');
it is seen to agree very well with the measured correlation.
\par
\vskip0.5cm
\noindent
{\bf 4.\ Nuclear Absorption and Anomalous Suppression}
\vskip0.5cm
With the relation between the measured transverse energy $E_T$ and the
impact parameter $b$ of the collision determined, we can now
calculate the $E_T$ dependence of the charmonium survival probability
in nuclear matter. It is given by
\ben
S_{Gl}(E_T) =
\int d^2b ~ P_{AB}(E_T,b)~
\left[ {d S_{Gl}(b) \over d^2 b} \right]~ \times
\nonumber
\een
\be
\left\{ \int d^2b ~ P_{AB}(E_T,b)~
\left[{d S_{Gl}(b;\sigma_{abs}=0) \over d^2b}\right] \right\}^{-1}.
\label{20}
\ee
Here the survival probability at fixed impact parameter is given by Eq.\
(\ref{6}) and the $E_T-b$ distribution by Eq.\ (\ref{14}). Through Eq.\
({\ref{20}), $S_{Gl}(E_T)$ is normalized
at fixed $E_T$ such that $S_{Gl}(E_T)=1$ for $\sigma_{abs}=0$.
With $\sigma_{abs}=7.3 \pm 0.6$ mb, from Eq.\ (\ref{4}), and with
$P_{AB}(E_T,b)$ as determined in the last section, we have absolute
predictions for the nuclear absorption suffered by charmonia in
nucleus-nucleus collisions.
\par
We begin with \J~production and first consider the $E_T$-integrated
survival probabilities in $O-Cu$, $O-U$ and $S-U$ collisions. In this
case, the experimental survival probabilities are given by
$S_{\psi}^{exp} = (\sigma^{exp}_{AB}/AB)/(\sigma^{exp}_{pD}/2)$ and
thus
fully determined in terms of measured cross sections. As seen in Table
2, the Glauber predictions fall well within the error range of the data.

\begin{table}
\begin{center}
\begin{tabular}{|c||c||c||c|}
\hline
& & & \\
& $S_{\psi}^{exp}$ & $S_{Gl}$ & $S_{\j}^{corr}$ \\
& & (7.3$\pm$0.6 mb) & (7.3$\pm$0.6 mb) \\
& & &\\
\hline
\hline
O-Cu & 0.57 $\pm$ 0.06 & 0.63 $\pm$ 0.02 & ~~\\
\hline
O-U  & 0.53 $\pm$ 0.05 & 0.52 $\pm$ 0.03 & ~~ \\
\hline
S-U  & 0.46 $\pm$ 0.05 & 0.49 $\pm$ 0.03 & 0.47 $\pm$ 0.03\\ \hline
\hline Pb-Pb & 0.27 $\pm$ 0.02 & 0.39 $\pm$ 0.03 & 0.37 $\pm$ 0.03\\
\hline
\end{tabular}\end{center}

\caption{$J/\psi$ suppression in $A-B$ collisions [2,10].}
\end{table}

\par
Turning now to the $E_T$-dependence, we list in Table 3a the
experimental survival probabilites \cite{Carlos,Borhani}, obtained
with reference to the rescaled $p-D$ data of NA51. They are shown
together with the corresponding Glauber results, which again agree well
with the data.
\par

\begin{table}

\begin{center}
\begin{tabular}{|c||c||c||c|}
\hline  & & &   \\
$<E_T>$ & $S_{\psi}^{exp}(E_T)$ & $S_{Gl}(E_T)$ &
$S_{\j}^{corr}(E_T)$ \\
GeV & &(7.3$\pm$0.6 mb)  &  (7.3$\pm$0.6 mb)  \\
& & &\\
\hline \hline
25 & 0.54 $\pm$ 0.05 & 0.58 $\pm$ 0.02  & 0.56 $\pm$ 0.03 \\
\hline
42 & 0.49 $\pm$ 0.05 & 0.52 $\pm$ 0.03  & 0.49 $\pm$ 0.03 \\
\hline
57 & 0.45 $\pm$ 0.05 & 0.48 $\pm$ 0.03  & 0.46 $\pm$ 0.03 \\
\hline
71 & 0.43 $\pm$ 0.05 & 0.45 $\pm$ 0.03  & 0.43 $\pm$ 0.03 \\
\hline
82 & 0.42 $\pm$ 0.04 & 0.43 $\pm$ 0.03  & 0.41 $\pm$ 0.03 \\
\hline
\end{tabular}
\vskip 0.8cm
(a) $J/\psi$ suppression in S-U collisions
\vskip 1.2cm
\begin{tabular}{|c||c||c|c|}
\hline  & &  \\
$<E_T>$ & $S_{\psi'}^{exp}(E_T)$ & $S_{Gl}(E_T)$  \\
GeV& & (7.3$\pm$0.6 mb)  \\
 & &\\
\hline \hline
25 & 0.35 $\pm$ 0.07 & 0.58 $\pm$ 0.02  \\ \hline
42 & 0.25 $\pm$ 0.05 & 0.52 $\pm$ 0.03  \\ \hline
57 & 0.19 $\pm$ 0.04 & 0.48 $\pm$ 0.03  \\ \hline
71 & 0.13 $\pm$ 0.03 & 0.45 $\pm$ 0.03  \\ \hline
82 & 0.09 $\pm$ 0.02 & 0.43 $\pm$ 0.03  \\ \hline
\end{tabular}
\vskip 0.8cm
(b) $\psi'$ suppression in S-U collisions
\end{center}
\vskip 0.5cm
\caption{Charmonium suppression in S-U collisions [NA38].}

\end{table}

\begin{table}

\begin{center}
\begin{tabular}{|c||c||c||c|}
\hline  & &   \\
$<E_T>$ & $ [(\j)/DY]_{exp}(E_T)$ & $ 46.0~S_{\j}^{corr}(E_T)$
\\
GeV & &(7.3$\pm$0.6 mb) \\
 & &\\
\hline \hline
25 & 25.2 $\pm$ 0.7 & 25.7 $\pm$ 1.2 \\
\hline
42 & 22.8 $\pm$ 0.6 & 22.8 $\pm$ 1.2  \\
\hline
57 & 21.0 $\pm$ 0.5 & 21.0 $\pm$ 1.2  \\
\hline
71 & 20.2 $\pm$ 0.4 & 19.6 $\pm$ 1.2   \\
\hline
82 & 19.2 $\pm$ 0.4 & 18.7 $\pm$ 1.2   \\
\hline
\end{tabular}
\vskip 0.8cm
(a) $S-U$ collisions [NA38]

\vskip 1.2cm

\begin{tabular}{|c||c||c||c|}
\hline  & &   \\
$<E_T>$ & $ [(\j)/DY]_{exp}(E_T)$ & $ 46.0~S_{\j}^{corr}(E_T)$
\\
GeV & &(7.3$\pm$0.6 mb) \\
 & &\\
\hline \hline
35 & 17.8 $\pm$ 2.2 & 20.2 $\pm$ 1.2 \\
\hline
59 & 13.2 $\pm$ 1.0 & 17.9 $\pm$ 1.2  \\
\hline
88 & 12.7 $\pm$ 0.8 & 16.6 $\pm$ 1.2  \\
\hline
120 & 11.4 $\pm$ 0.8 & 15.6 $\pm$ 1.2   \\
\hline
149 & 8.6 $\pm$ 0.8 & 15.2 $\pm$ 1.2  \\
\hline
\end{tabular}
\vskip 0.8cm
(b) $Pb-Pb$ collisions [NA50]
\end{center}
\vskip 0.5cm
\caption{($J/\psi$)/Drell-Yan ratios in nuclear collisions.}
\end{table}

\begin{table}
\begin{center}
\begin{tabular}{|c||c||c||c|c|}
\hline  & & & \\
$<E_T>$ & $S_{\psi}^{exp}(E_T)$ & $S_{Gl}(E_T)$  &
$S_{\j}^{corr}(E_T)$ \\
GeV& &(7.3$\pm$0.6 mb) &  (7.3$\pm$0.6 mb)  \\
&  & &  \\
\hline \hline
35 & 0.40 $\pm$ 0.06 & 0.47 $\pm$ 0.03 & 0.44 $\pm$ 0.03 \\
\hline
59 & 0.30 $\pm$ 0.03 & 0.41 $\pm$ 0.03 & 0.39 $\pm$ 0.03 \\
\hline
88 & 0.29 $\pm$ 0.03 & 0.38 $\pm$ 0.03 & 0.36 $\pm$ 0.03  \\
\hline
120 & 0.26 $\pm$ 0.03 & 0.36 $\pm$ 0.03 & 0.34 $\pm$ 0.03  \\
\hline
149 & 0.20 $\pm$ 0.02 & 0.35 $\pm$ 0.03 & 0.33 $\pm$ 0.03  \\
\hline
\end{tabular}
\vskip 0.8cm
(a) $J/\psi$ suppression in Pb-Pb collisions
\vskip 1.2cm

\begin{tabular}{|c||c||c|c|}
\hline  & &  \\
$<E_T>$ & $S_{\psi'}^{exp}(E_T)$ & $S_{Gl}(E_T)$  \\
GeV& fm & (7.3$\pm$0.6 mb)  \\
 & & \\
\hline \hline
51 & 0.12 $\pm$ 0.02 & 0.43 $\pm$ 0.03  \\ \hline
84 & 0.11 $\pm$ 0.02 & 0.38 $\pm$ 0.03  \\ \hline
115 & 0.08 $\pm$ 0.02 & 0.36 $\pm$ 0.03  \\ \hline
144 & 0.06 $\pm$ 0.01 & 0.35 $\pm$ 0.03  \\ \hline
\end{tabular}
\vskip 0.8cm
(b) $\psi'$ suppression in Pb-Pb collisions
\end{center}
\vskip0.5cm
\caption{Charmonium suppression in Pb-Pb collisions [NA50].}
\end{table}

\par
We thus find that the \J~suppression observed in
the collision of $O$ and $S$ beams with nuclear targets can be
accounted for by pre-resonance absorption in nuclear matter.
For \P~production, this is no longer the case. The data for the
survival probabilities are shown in
Table 3b. They are well below the nuclear absorption predictions, and
the additional suppression increases with increasing $E_T$. We shall
return in Section 5 to the interpretation and description
of this enhanced \P~suppression; here we only want to note its effect
on \J~production. It is known that $8\pm 2$ \% of the observed \J's are
due to \P~decay \cite{HP}. Since the latter is suppressed in $S-U$ collisions,
the corresponding fraction of the observed \J's must be suppressed
as well. We thus obtain as corrected \J~suppression
\be
S_{\j}^{corr} = 0.92~S_{Gl} + 0.08~S_{\psi'} ; \label{21}
\ee
the corresponding values are included in the Tables and are seen to
generally bring data and nuclear absorption results even closer
together.
\par
As a further check with data of higher precision, we consider the
directly measured ratio of \J~to Drell-Yan production. The
$E_T$-dependence of this ratio is predicted as
\be
[\sigma_{\psi}^{AB}/\sigma_{DY}^{AB}](E_T) =
G~S_{\j}^{corr}(E_T),
\label{22}
\ee
with $G = \sigma_{\psi}^{NN}/\sigma_{DY}^{NN}$. Since the
nucleon-nucleon ratio at 200 GeV/c beam momentum is
experimentally not determined with the same precision as the nuclear
ratios, we fit the form (\ref{22}) to the data
\cite{Carlos,Borhani} and check
whether the resulting $E_T$ dependence is correct. The result, with
$G=46.0$, is shown in Table 4a and seen to give excellent agreement
with the data.
\par
Next we turn to the new data for \J~production in $Pb-Pb$ collisions.
The integrated \J~survival probability as predicted by Glauber theory,
with the pre-resonance nuclear absorption cross section (\ref{4}), is
included in Table 2, together with the corresponding data; 
data and prediction for the $E_T$-dependence
are listed in Tables 4b and 5a. We note that the data decrease well
below the predictions: while the survival probability measured at the
lowest $E_T$ value is close to pre-resonance absorption,
that at the highest $E_T$ is about 40\% smaller. We thus have a
clear and $E_T$-dependent onset of additional \J~suppression.
\par
The essential result of this section is thus  
summarized in Fig.\ 6: for $S-U$ collisions, nuclear 
absorption fully accounts for the observed attenuation, for $Pb-Pb$ collisions
there is an additional and in this sense ``anomalous'' suppression.
\par
For the \P, an enhanced suppression had already been seen in $S-U$
collisions. In $Pb-Pb$ interactions, the suppression beyond nuclear
absorption increases even more, as shown in Table 5b. It also continues
to become stronger with increasing $E_T$.
\vskip0.5cm
\noindent
{\bf 5.\ Suppression by Hadronic Comovers}
\vskip0.5cm
Before we can use charmonium suppression as probe for deconfinement, we
have to determine to what extent suppression beyond nuclear absorption
can be accounted for by dissociation in a ``normal" confined medium.
Consider a charmonium state formed at the center of the two colliding
nuclei at the moment of their complete overlap, at rest in the center of
mass, i.e., at $x_F=0$. For the SPS c.m.s. energy of 17 -- 20 GeV, the
nucleon distribution within each nucleus is Lorentz-contracted along the
beam axis to a disc of some 1 -- 2 fm thickness. This implies that after
about 0.5 -- 1.0 fm, most of the nucleons have swept over the nascent
charmonium state, resulting in the nuclear absorption described above.
After this time, it finds itself within the medium produced by the
collision, and it is this medium we want to probe.
\par
The initial density of hadronic comovers is in general determined by
the density of wounded nucleons \cite{Albrecht}. Since the integrand
$[dN_w(b,s)/d^2s]$ of Eq.\ (\ref{9}) is the local density of wounded
nucleons at fixed impact parameter, the average density associated to
hard collisions is given by
\be
n_w(b) = \int d^2s ~\left[ {dN_c \over d^2s}(b,s) \right]
\left[ {dN_w\over d^2s}(b,s)\right]  \Big/  \int d^2s
\left[{dN_c \over d^2s}(b,s)\right]  .\label{25}
\ee
Convolution with $P_{AB}(E_T,b)$ will convert expression (\ref{25})
into the corresponding density at fixed $E_T$; the resulting behaviour
is shown in Fig.\ 7.
\par
Because of the very low binding energy ($\sim$ 60 MeV), the \P~could
readily be dissociated by hadronic comovers. To study this, we first
express the initial comover density in terms of the wounded nucleon
density,
\be
n_{co}(b,s) = N_h~\left[ {dN_w(b,s) \over d^2s } \right], \label{26}
\ee
where $N_h$ is the number of comoving hadrons produced by each wounded
nucleon. We then have to take into account that comover absorption
occurs after the absorption of the pre-resonance state; in other
words, only a \P~which has survived the pre-resonance passage through
the nuclear medium can be absorbed by comovers. The survival
probability of a \P~in a comover medium undergoing isentropic
longitudinal expansion therefore becomes
\ben
S^{co}_{\psi'} (b) = \int d^2s~
\exp\{- v \sigma_{h\psi'} N_h [dN_w(b,s)/d^2s]
\log\{N_h[dN_w(b,s)/d^2s]/n_f\}\} ~\times
\een
\be
\left\{{d^2 S_{Gl}\over d^2b~ d^2s}(b,s)\right\}~
\Bigg/~ \left\{ \int d^2s {d^2 S_{Gl} \over
d^2s~d^2b}(b,s;\sigma_{abs}=0) \right\},
\label{27}
\ee
where $\sigma_{h\psi'}$ denotes the cross section for \P~break-up by
collisions with hadronic comovers, $v$ the average relative velocity
between the \P~and the colliding hadrons, and $n_f$ the (universal)
freeze-out density for comovers. In Eq.\ (\ref{27}) we have assumed
the cross section for \P~dissociation by hadrons to be
energy-independent; since $2 M_D -M_{\psi'} \simeq 60$ MeV,
$\sigma_{h\psi'}$ is expected to attain its geometric value of about
10 mb very near threshold. Rather than estimate the various quantities
involved, we prefer to use the \P~suppression in $S-U$ collisions to
determine the two constants $c_1 \equiv N_h/n_f$ and $c_2\equiv v
\sigma_{h\psi'} N_h$ empirically and then verify that the resulting
values are of the right magnitude. In Fig.\ 8a we see that with
$c_1=1.3~{\rm fm}^2$ and $c_2=0.8$, such a comover description indeed
agrees quite well with the \P~suppression observed in $S-U$
collisions. The density of wounded nucleons in a nucleon-nucleon
collision is about 1 fm$^{-2}$; taking this as $n_f$, and using
$N_h\simeq 1.5$, we would get $c_1 \simeq 1.5$.
For a $h-\psi'$ cross section of 10 mb, $c_2=0.8$
fm$^{-2}$ implies a hadron velocity of about 0.5. All constants thus
lead to reasonable values of the involved quantities.
\par
To compare this comover absorption to the
(lower energy) $Pb-Pb$ data, we rescale $c_1$ and $c_2$ by the factor
1.4/1.5 found above in minimum bias collisions. In Fig.\ 8b we see
that the result agrees also with the $Pb-Pb$ data. Finally,
in Fig.\ 8c, we show the combined $S-U$ and $Pb-Pb$ data as function of
the average comover density $n_{co}(E_T) = N_h~n_w(E_T)$ in
charmonium production; the observed behaviour scales quite well in
$n_{co}$.
\par
The extension of such a comover approach to anomalous \J~suppression is
confronted with two problems. Heavy quark QCD calculations \cite{KS3}
exclude \J~break-up by interaction with hadrons in the present energy
range, since the dissociation cross section is strongly damped near
threshold; this remains true also when target mass effects are included
\cite{Gena}. However, it is not {\sl a priori}
clear that charm quarks are sufficiently heavy to assure the
applicability of heavy quark theory. Although this theory correctly
reproduces \J-photoproduction data, it can and certainly
should be checked directly \cite{KS5}. Until such a direct experimental
confirmation is given, it remains of interest to assume less or
no threshold damping; the \J~will then interact with a hadronic
comover medium through much or all of its geometric cross
section.
\par
The second problem is that the suppression by hadronic comovers
increases continuously; there cannot be a sudden onset. Hence it is
necessary to check if the presence of comover absorption in $S-U$
collisions is compatible with the constraints provided by all
presently available data.
The \J~absorption by hadronic comovers is readily obtained from Eq.\
(\ref{27}); we just replace the \P-hadron break-up cross section by
that for \J~dissociation. If we assume this cross section to be around
1.5 mb (i.e., approximately half its geometric value) essentially from
threshold on, we obtain as much suppression as is observed in $Pb-Pb$
collisions, as seen in Fig.\ 9a. However, with the break-up cross
section thus fixed, we now have a parameter-free prediction for the
$S-U$ suppression, and it is evident from Fig.\ 9a that this does not
agree with the data. Fig.\ 9a also illustrates quite clearly that a
{\it precise} study of normal vs.\ anomalous behaviour is needed to
confirm this conclusion.
\par
We therefore turn to the ratio of (\J)/(Drell-Yan) production in
$S-U$ and $Pb-Pb$ collisions (Table 4). In Fig.\ 9b we see that with the
absorption cross section fixed at 1.5 mb, the \J~suppression
remains in good agreement with the $Pb-Pb$ data; however, the
corresponding comover absorption clearly misses the $S-U$ data.
In other words: if we tune the \J-comover
dissociation cross section such as to obtain the amount of suppression
observed in $Pb-Pb$ collisions, then the only way the data have of
showing dissent is to disagree with the corresponding $S-U$ prediction;
and this is what they do.
\par
Finally we show in Fig.\ 10 the ratio of \J~to Drell-Yan production in
$S-U$ and $Pb-Pb$ collisions as function of $n_w$. One way to relate
the $Pb-Pb$ data with $n_w$ (filled triangles) is obtained by
applying our $b-E_T$ assignment to the measured $<\!E_T\!>$ points, using
the Woods-Saxon nuclear distribution. For a second way (open triangles), we
use the $b-E_T$ assignment obtained by NA50 \cite{Gonin} from the
$E_T$-distribution including the low $E_T$ detector deficiency and using
a hard core nuclear distribution. We note that the precise position of
the peripheral data points as function of $n_w$
is rather dependent on the details of the collision geometry and the
nuclear distribution. Hence here conclusive results may have to
wait for higher statistics and more complete low $E_T$ $Pb-Pb$ data.
\par
Let us now comment briefly on some proposed comover descriptions of
anomalous \J~suppression. The first such study \cite{G-V} including the
$Pb-Pb$ data postulates ad hoc a comover density different
from that in Glauber theory, setting
\be
n_{co}(E_T) = c ~E_T~, \label{28}
\ee
in terms of the hadronic transverse energy. This relation, although
correct in the fluctuation regime, is not  in accord with the
peripherality of the actually measured collisions. The comover
density as function of $E_Z$ is given directly by Glauber theory,
relating the number of unwounded to that of wounded nucleons. Combining
this with Eq.\ (\ref{28}) gives the $E_T-E_Z$ correlation provided
by ref.\ \cite{G-V}. It is included in Fig.\ 5 and seen to disagree
strongly with the measured form. The model thus does not reproduce the
actual collision conditions of the experiment. It also uses
$\sigma_{abs} = 4.8$ mb, which is not compatible with Eq.\ (\ref{4}).
\par
A subsequent, more detailed comover investigation \cite{C-G} uses Glauber
theory consistently throughout. To account for comover interactions, it
requires the corresponding cross sections for \J-hadron and \P-hadron
interactions; in addition, it takes into account $\j \to \psi'$ and
$\psi' \to \j$ transitions. Together with the pre-resonance absorption
cross section, one thus has a total of five adjustable parameters, to be
tuned such as to account for the anomalous \J~suppression in $Pb-Pb$
collisions as well as for the $S-U$ and $p-A$ data. Let us illustrate
what this leads to. Our (\P-corrected) nuclear absorption results in
Table 4a fit the data listed there with a $\chi^2/d.f.$ of about 1. The
\J~behaviour predicted in ref.\ \cite{C-G} leads to a $\chi^2/d.f.$ of
about 4 and is thus excluded with better than 99\% confidence level. In
our opinion, ref.\ \cite{C-G} thus establishes nicely that even after
the introduction of four further parameters, a comover picture can
reproduce the \J-suppression observed in $Pb-Pb$ collisions only at
the expense of giving up an acceptable account of the $S-U$ data.
\par
We therefore conclude that it is not possible to describe consistently
in terms of hadronic comovers both the ``normal" (pre-resonance)
\J~suppression observed in $S-U$ collisions and the anomalous
\J~suppression in $Pb-Pb$ interactions. In contrast, the measured
\P~suppression is well described by hadronic comover absorption.
\vskip0.5cm
\noindent
{\bf 6.\ Suppression by Colour Deconfinement}
\vskip0.5cm
For the fate of a \J~in dense matter, the basic difference between
comover absorption and colour deconfinement is that the former is always
present, changing smoothly with kinematic conditions, whereas the latter
has no effect below some critical threshold. For colour deconfinement,
the interaction region thus consists of a ``hot" (inner) part, where a
\J~can be dissociated, and ``cool" (outer) part leaving it intact. In
the case of small nuclei or very peripheral collisions, no region is
hot enough, and so there will be only pre-resonance nuclear absorption.
Hence the relative fraction of hot interaction region becomes the
determining variable for deconfinement suppression \cite{GS2}.
\par
As already noted, the \J~plays a crucial role in probing colour
deconfinement by charmonium suppression: while the weakly bound \P~is
easily broken up in confined as well as in deconfined media, only hard
and hence deconfined gluons can resolve the structure of the small
\J~and overcome its binding energy $2M_D - M_{\psi} \simeq 0.64$ GeV
\cite{KS3}. We therefore concentrate here on \J~suppression. Since it is
not clear from Fig.\ 9 if the density of wounded nucleons (and hence
also the initial energy density) is a suitable variable to describe the
observed \J~suppression pattern, we do not want to assume an
equilibrated system.
\par
The precursor of a quark-gluon plasma is a system in which the partonic
constituents are no longer distributed in the way they are in hadrons.
A change in the gluon distribution can arise by interactions between
wounded nucleons. The amount of internetting between wounded nucleons
is thus crucial: how many collisions does a wounded
nucleon undergo? This is measured by the interaction density
\cite{K-Heidel}
\be
\kappa \equiv {N_c \over N_w} \geq 0.5, \label{29}
\ee
where $N_c$ and $N_w$ denote the number of nucleon-nucleon collisions
and that of wounded nucleons, respectively. Since in $p-A$ collisions
$N_w=N_c+1$, $\kappa$ here lies in the range $0.5 \leq \kappa \leq 1$,
with $\kappa=0.5$ for $p-p$ interactions. It can become larger than
unity only in $A-B$ collisions, since there interactions between wounded
nucleons are possible, and it increases with the centrality of the
collisions. Since Glauber theory provides both $N_w$ and $N_c$,
$\kappa$ is readily calculable.
\par
The interaction measure $\kappa$ is of particular interest because it
can be determined in a model-independent way from experimental data.
The number $N_{DY}$ of Drell-Yan pairs produced in an $A-B$
interaction is directly proportional to the total number of
nucleon-nucleon collisions, while the number $N_h$ of produced
hadronic secondaries (or the transverse hadronic energy
$E_T$) is determined by the number of wounded nucleons. We thus have
$\kappa \sim N_{DY}/N_h$ in terms of measurable quantities
and can cross-correlate charmonium suppression at a
given $E_T$ (and hence at a given degree of internetting) with other
observables which might show a rescattering dependence, such as ratios
of hadronic secondaries.
\par
For the onset of deconfinement, we assume that once $\kappa$
reaches some critical value $\kappa_c$, there are enough interactions
between wounded nucleons to provide the hard gluons necessary for
\J~dissociation. Consider $\kappa(b,s)$ at fixed impact parameter $b$,
as function of the profile $s$ of the interaction region.
The condition $\kappa(b,s_c)=\kappa_c$ fixes the extension $s_c$ of the
hot volume at the given $b$. We now evaluate the \J~survival
probability (\ref{6}) with the constraint that the integrand vanishes
for $s \leq s_c$. The resulting $S_{\j}^{dec}(b)$ we convolute with
$P(b,E_T)$ to obtain the predicted survival probability at fixed $E_T$.
\par
Before we can compare this result to the data, we have to take into
account that approximately 40\% of the produced \J's come from
$\chi_c$ decay (see ref.\ \cite{HP} for details),
and the $\chi_c$, with its larger radius
($r_{\chi}\simeq 0.4~{\rm fm} > r_{\psi} \simeq 0.2$ fm) and smaller
binding energy ($\Delta E_{\chi} \simeq 0.3~{\rm GeV} < \Delta E_{\psi}
\simeq 0.6$ GeV), is more easily dissociated than the \J.
It is not clear if $\chi$-hadron interactions can be treated by heavy
quark QCD in the same way as those for \J's; however, the formalism
does give reasonable results for the $\chi$ as well \cite{KS5}. We
thus assume that there are two distinct $\kappa_c$ values:
$\kappa_c^{\chi}$ and $\kappa_c^{\psi} > \kappa_c^{\chi}$. Since the
$S-U$ data show no anomalous suppression, $\kappa_c^{\chi}$ has to be at
least as large as the maximum value of $\kappa(b,s)$ in
$S-U$ collisions. From
\be
\kappa_c^{\chi} = \left[ \kappa(b=0,s=0) \right]_{SU} \simeq 2.3
\label{30}
\ee
we thus get a possible starting point for anomalous suppression. The
value of $\kappa_c^{\psi}$ is considered as open parameter and
determined
such as to get best agreement with the data. The overall \J~survival
probability $S_{\j}^{dec}(E_T)$ is now obtained by combining 40\%
$\chi$-suppression with 60\% $\psi$-suppression, the latter reduced by
the suppressed \P~component. In Fig.\ 11 we see that with
$\kappa_c^{\psi}=2.9$, we do in fact get quite good agreement
with the results of both Table 4 and 5.
\par
The natural variable determining the amount of suppression
in this picture is the fraction $f_{\j}$ of \J's inside the hot
$\kappa$-region, compared to the overall interaction volume. It is
given by
\be
f_{\j}(b) = \int_0^{s_c} d^2 s~N_c(b,s)~S_{Gl}(b,s) \Bigg/
\int_0^{\infty} d^2 s~N_c(b,s)~S_{Gl}(b,s) \label{31}
\ee
The \J~survival probability then becomes
\be
S_{\j}^{dec}(b)/S_{Gl}(b) = [1 - f_{\j}(b)]; \label{32}
\ee
again the convolution with $P(E_T,b)$ provides the corresponding
expressions as function of $E_T$. In Fig.\ 12 we compare the data as
function of $f_{\j}$ with the prediction (\ref{32}); the agreement is
quite good.
\par
In \cite{Blaizot}, a similar approach is presented with the initial
energy density instead of $\kappa$ as relevant variable. Here also 
it is the relative amount of hot interaction region which determines
the suppression pattern. An interesting alternative starts deconfinement
at the percolation point of transverse string areas \cite{Pajares}.
\par
In summary, the observed anomalous \J~suppression pattern in $Pb-Pb$
collisions agrees quite well with a deconfinement interpretation.
\par
\vskip0.5cm
\noindent
{\bf 7.\ Conclusions}
\vskip0.5cm
Here we want to summarize our conclusions and list the further
experiments which we believe are needed to corroborate them.
\begin{itemize}
\item{Present $p-A$ data determine the cross section for pre-resonance
charmonium absorption in nuclear matter as $\sigma_{abs}=7.3 \pm 0.6$
mb.}
\item{The \J~suppression in $O-Cu$, $O-U$ and $S-U$ collisions is
fully accounted for by the corresponding nuclear absorption;
in contrast, $Pb-Pb$ collisions show additional (``anomalous") suppression
increasing with centrality (see Fig.\ 6).}
\item{\P~production is suppressed beyond nuclear
absorption already in $S-U$ collisions; this increases further in
$Pb-Pb$ collisions.}
\item{Anomalous \J~suppression present only in $Pb-Pb$ collisions
rules out any explanation in terms of hadronic comovers; all proposed
models are inconsistent with some of the available data.}
\item{Deconfinement, with an onset at the interaction density of
central $S-U$ collisions, is in accord with all data.}
\end{itemize}
\par\noindent
In two steps of the analysis, further data are clearly necessary for
definitive conclusions.
\begin{itemize}
\item{So far, $S-U$ collisions provide normal, $Pb-Pb$ collisions
anomalous suppression. Data from an intermediate $A-A$ experiment
straddling the critical divide is needed to check the onset and help
in specifying the relevant scaling variable. A $Pb-Pb$ experiment at
lower energy (around 100 GeV/c beam momentum) would provide an
excellent cross check, if sufficient statistics can be obtained. }
\end{itemize}
\par\noindent
The confirmation of a well-defined onset would establish critical
behaviour. It is then necessary to confirm that this is colour
deconfinement.
\begin{itemize}
\item{According to heavy quark QCD calculations, quarkonium dissociation
is effectively not possible in a confined medium of accessible
temperatures. To check this result for charmonium,
a direct study of \J-nucleon interactions is needed (``inverse
kinematics" experiment \cite{KS5}).}
\end{itemize}
\par\noindent
Besides these two really fundamental questions, there are several other
points where further data could help considerably in underpinning the
conclusions ob\-tained here. Higher precision $p-A$ data at 450 or 200
GeV/c would reduce the error margin in the nuclear absorption cross
section and thus result in more precise Glauber theory predictions. If
these data provide both \J~and \P~cross sections, also the
$A$-independence of the \P/(\J) ratio can be confirmed with still
greater precision, thus corroborating pre-resonance absorption as the
underlying mechanism. Finally, a $Pb-U$ experiment could extend the
range of $\kappa$ (or of the relevant energy density) to higher values
and thus might provide a first check of a two-stage $\chi-\psi$
suppression pattern.
\newpage
\noindent
\centerline{\bf Acknowledgements}
\vskip0.5cm
We thank J.-P. Blaizot, B. Chaurand, L. Kluberg, J. Schu\-kraft
and E. Scomparin for helpful discussions.
\vskip 1cm

\renewcommand{\section}{\subsection}

\newpage

\begin{center}\mbox{
\epsfig{file=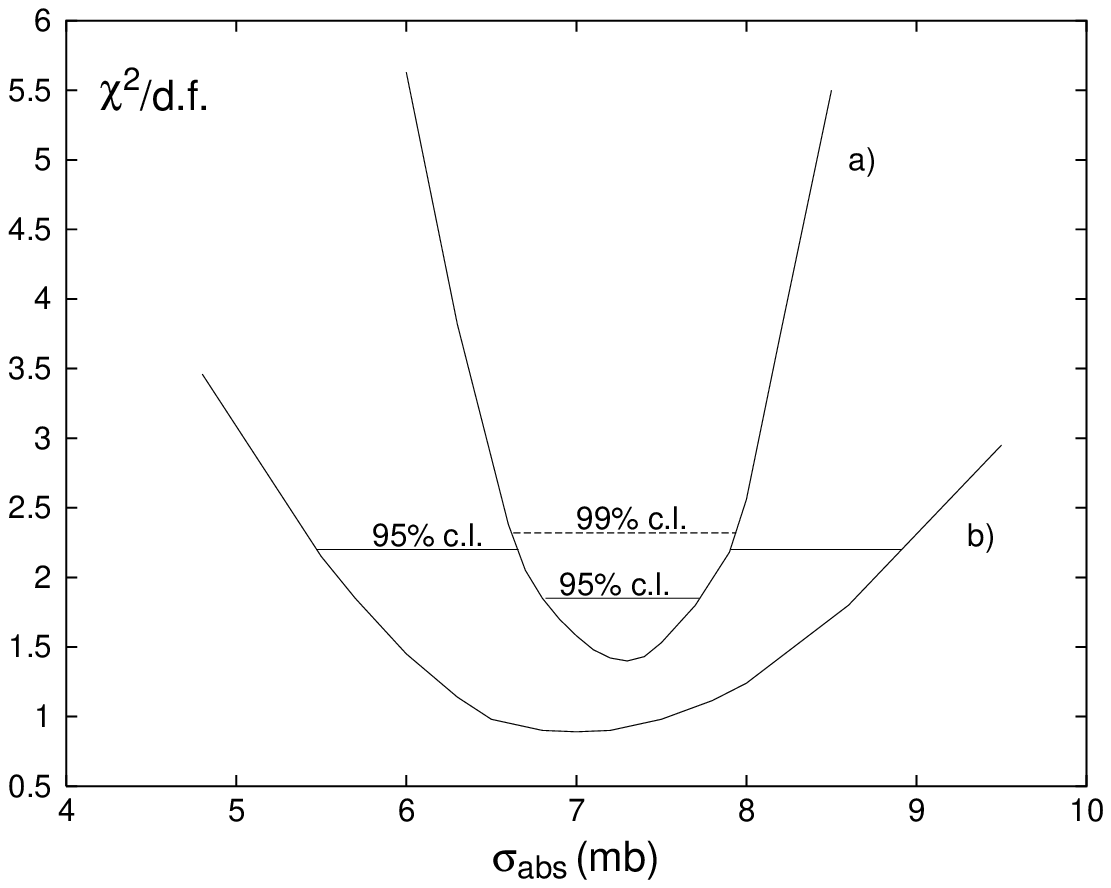,height=8cm}}\end{center}
Figure 1: ${\chi}^2$-distributions for nuclear cross section fits: a) relative
to $pD$, b)~relative to $pC$.
\begin{center}\mbox{
\epsfig{file=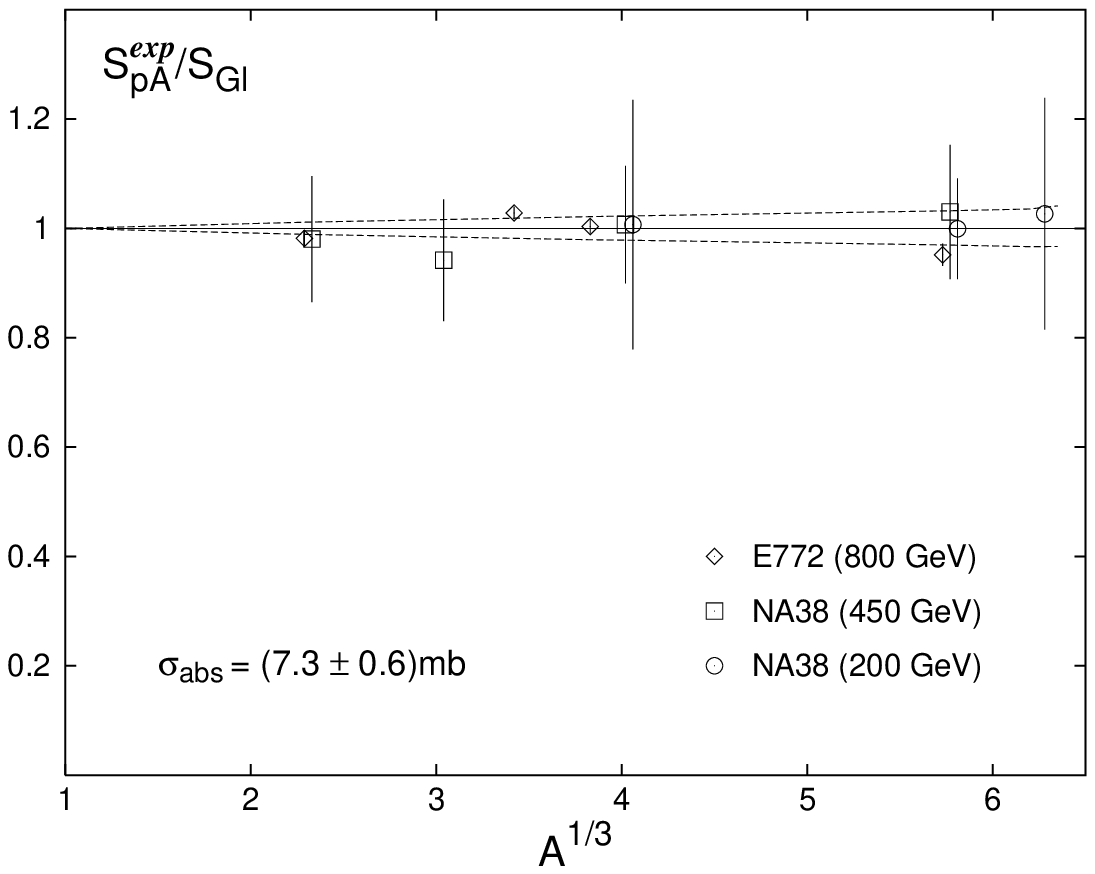,height=8cm}}\end{center}
Figure 2: $J/\psi$ survival probabilities in $p-A$ collisions, compared to
Glauber theory results with $\sigma_{abs}=7.3\pm 0.6$ mb.
\newpage
\vspace{-1.5cm}
\begin{center}\mbox{
\epsfig{file=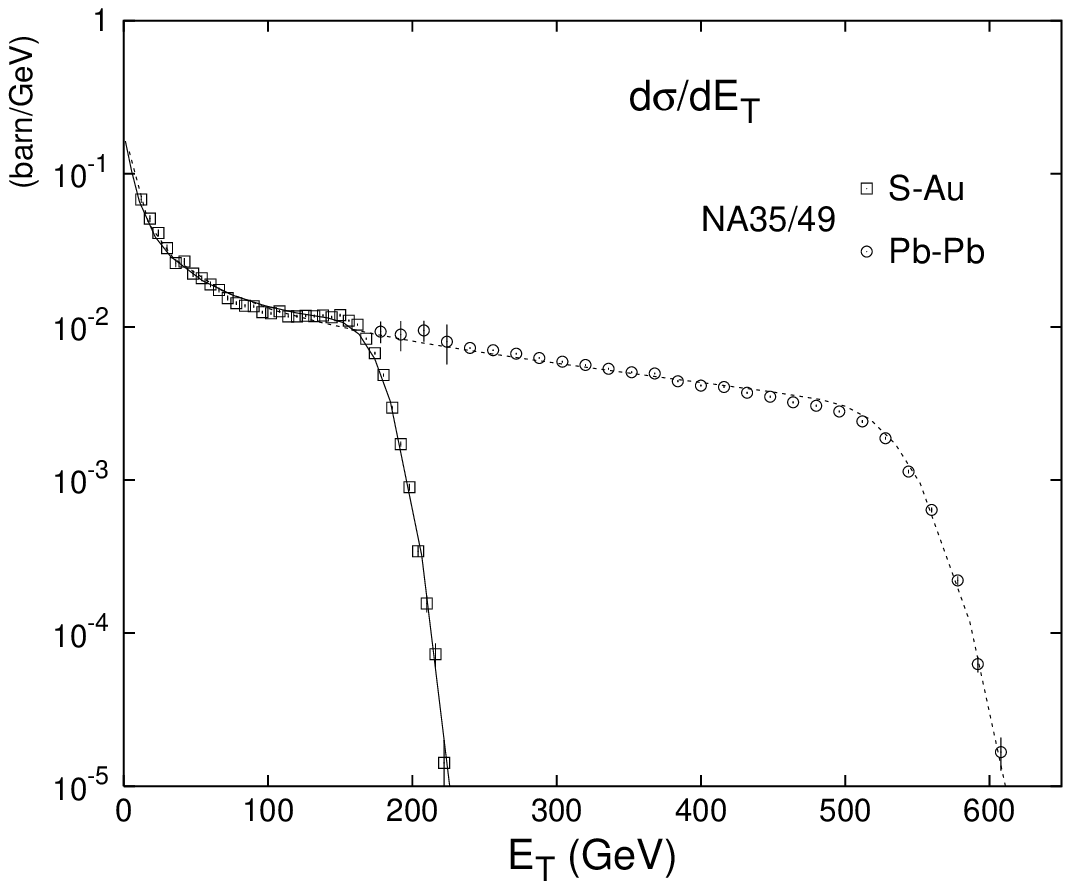,height=7.5cm}}
\end{center}
Figure 3: Minimum bias $E_T$ distributions for $S-Au$ and \mbox{$Pb-Pb$}
interactions \cite{Stock}, compared to Glauber theory results with
$q=1.5~(S-Au)$ and \mbox{$1.4~(Pb-Pb)$}.
\begin{center}
\mbox{
\epsfig{file=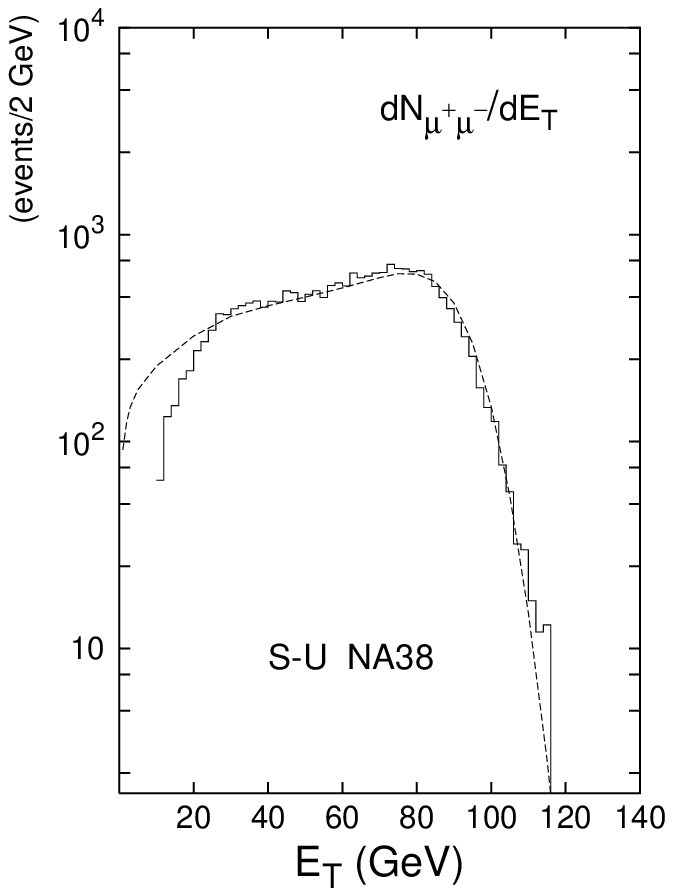,height=8cm}
\epsfig{file=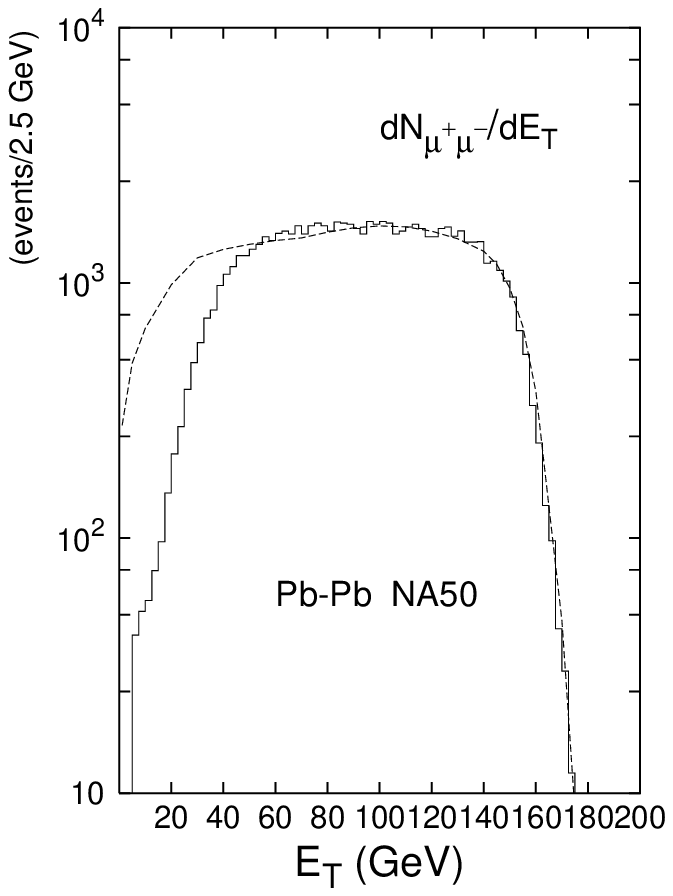,height=8cm}}\end{center}
Figure 4: Hard collision $E_T$ distributions for $S-U$
\cite{Carlos-Dis} and $Pb-Pb$ \cite{Gonin} interactions, compared to
Glauber theory results with $q=0.75~(S-U)$ and $0.40~(Pb-Pb)$.
\newpage
\vspace{-2.cm}
\begin{center}\mbox{\hspace{-.9cm}
\epsfig{file=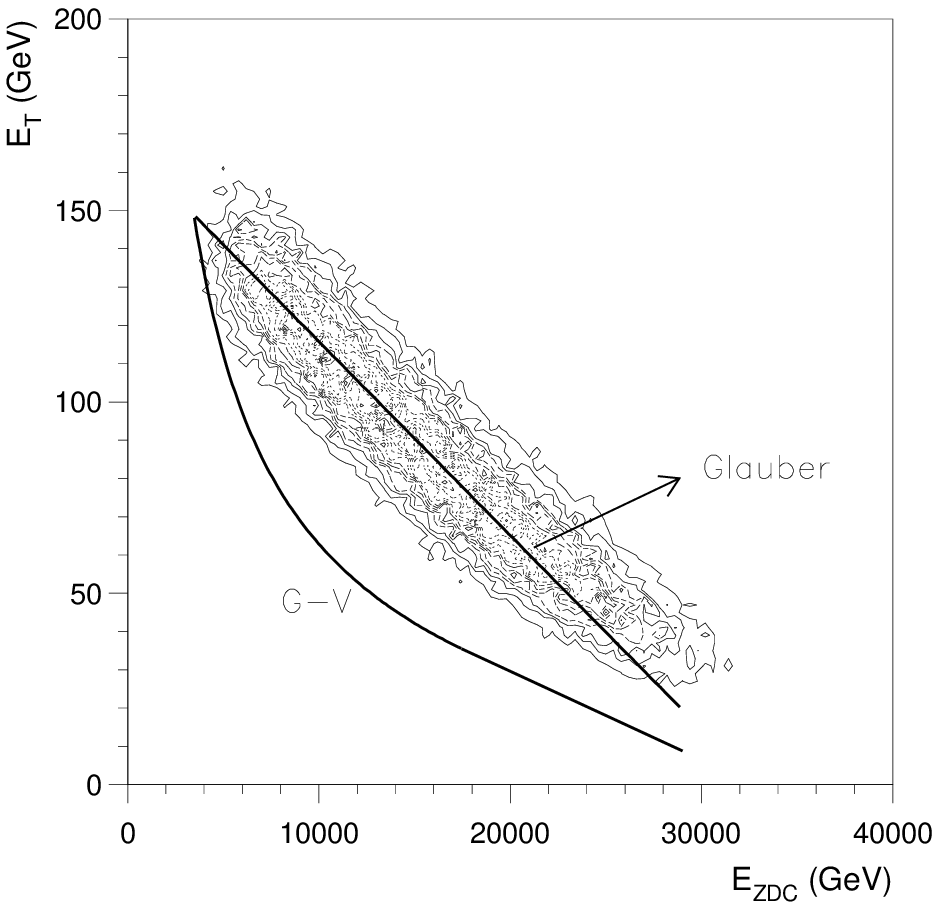,height=11cm}}\end{center}
Figure 5: $E_T-E_Z$ correlation in $Pb-Pb$ collisions
\cite{Scomparin}, compared to the Glauber theory correlation. The curve
labelled `G-V' is the correlation of model \cite{G-V}; see section 5.
\newpage
\begin{center}\mbox{
\begin{array}[b]{c}\left.\mathrm{a}\right) 
\\ \\ \\ \\ \\ \end{array}
\hspace{.6cm}
\epsfig{file=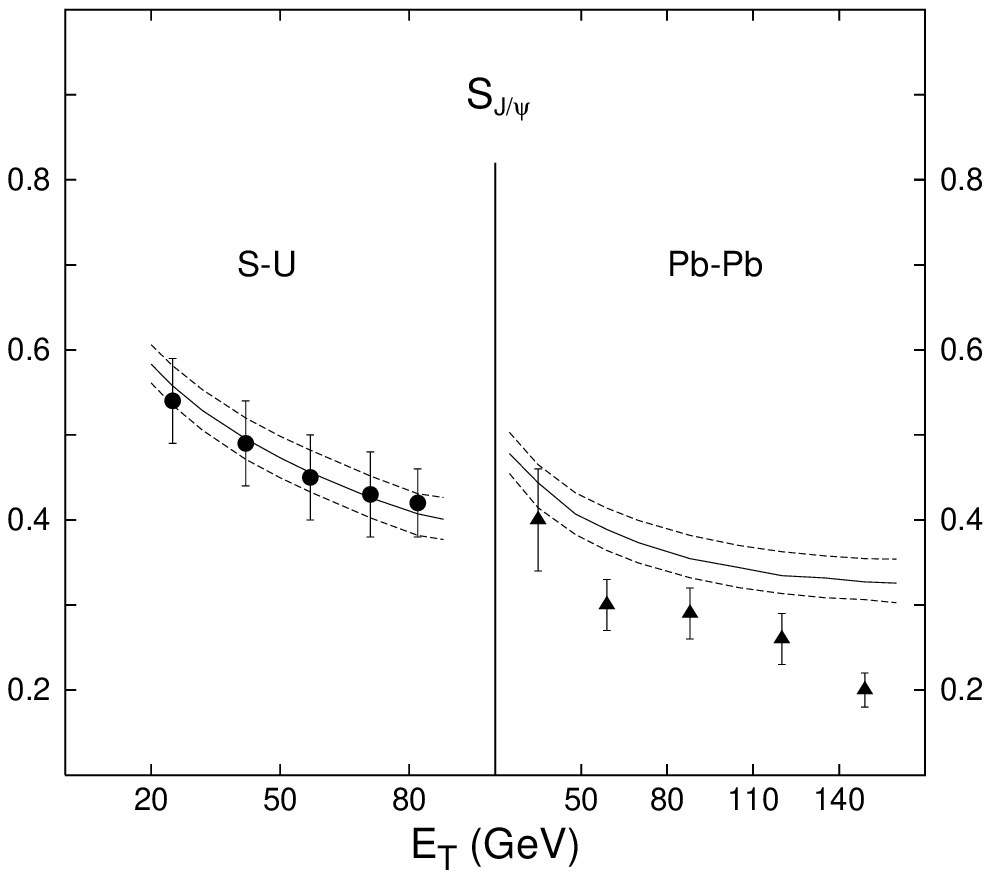,height=8cm}
\hspace{1.cm}
\begin{array}[b]{c}\left.\mathrm{ }\right.
\\ \\ \\ \\ \\ \end{array}
}\end{center}
\begin{center}\mbox{
\begin{array}[b]{c}\left.\mathrm{b}\right) 
\\ \\ \\ \\ \\ \end{array}
\hspace{.6cm}
\epsfig{file=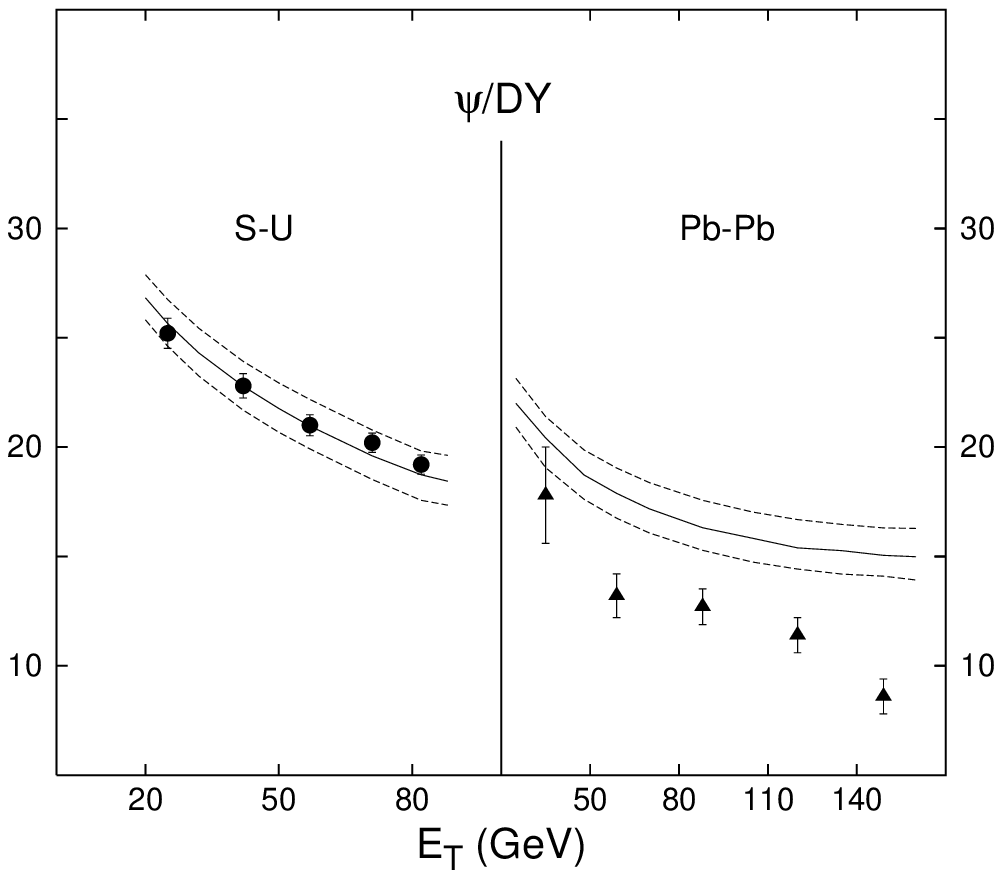,height=8cm}
\hspace{1.cm}
\begin{array}[b]{c}\left.\mathrm{ }\right.
\\ \\ \\ \\ \\ \end{array}
}\end{center}
Figure 6: $J/\psi$~suppression in $S-U$ and $Pb-Pb$ collisions for (a) the
survival probability and (b) the ratio 
of $J/\psi$~to Drell-Yan production.  The curves show the
Glauber theory results with $\sigma_{abs}=7.3\pm 0.6$ mb.
\newpage
\begin{center}\mbox{
\epsfig{file=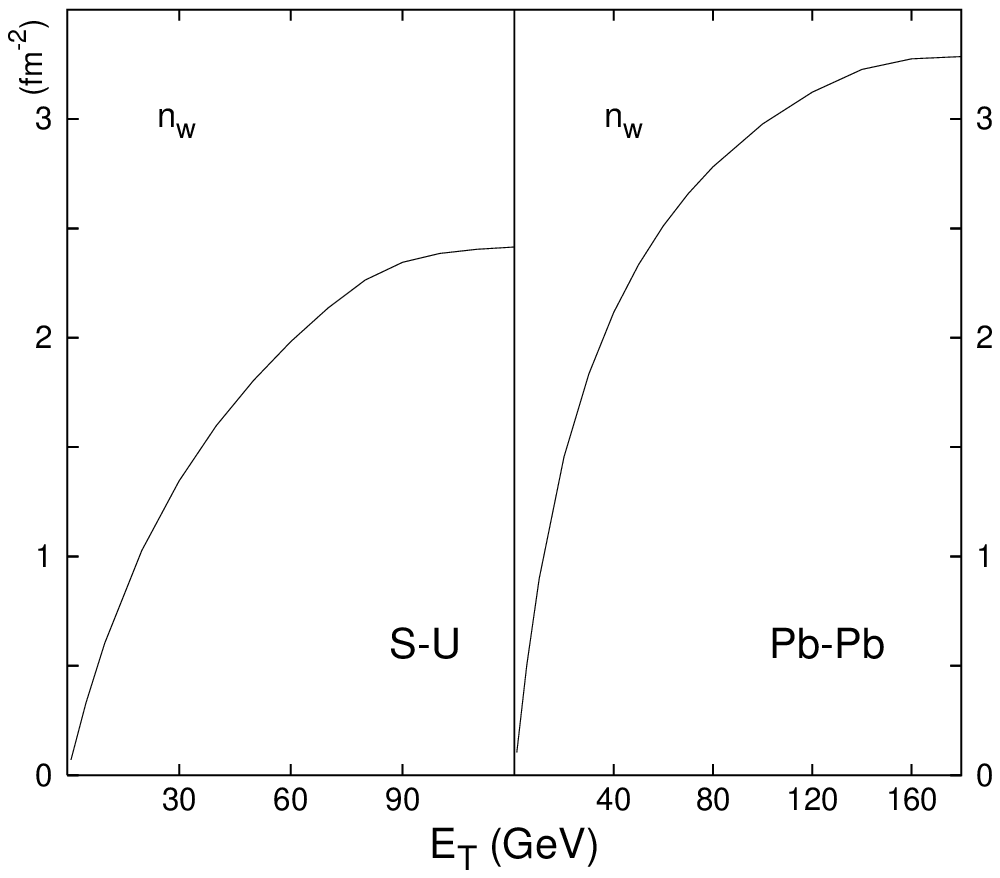,height=10cm}}\end{center}
Figure 7: The average density of wounded nucleons in 
hard interactions for $S-U$ and $Pb-Pb$ collisions.
\vspace{-.5cm}
\newpage
\begin{center}\mbox{\begin{array}[b]{c}\left.\mathrm{a}\right) 
\\ \\ \\ \\ \\ \end{array}
\hspace{.6cm}
\epsfig{file=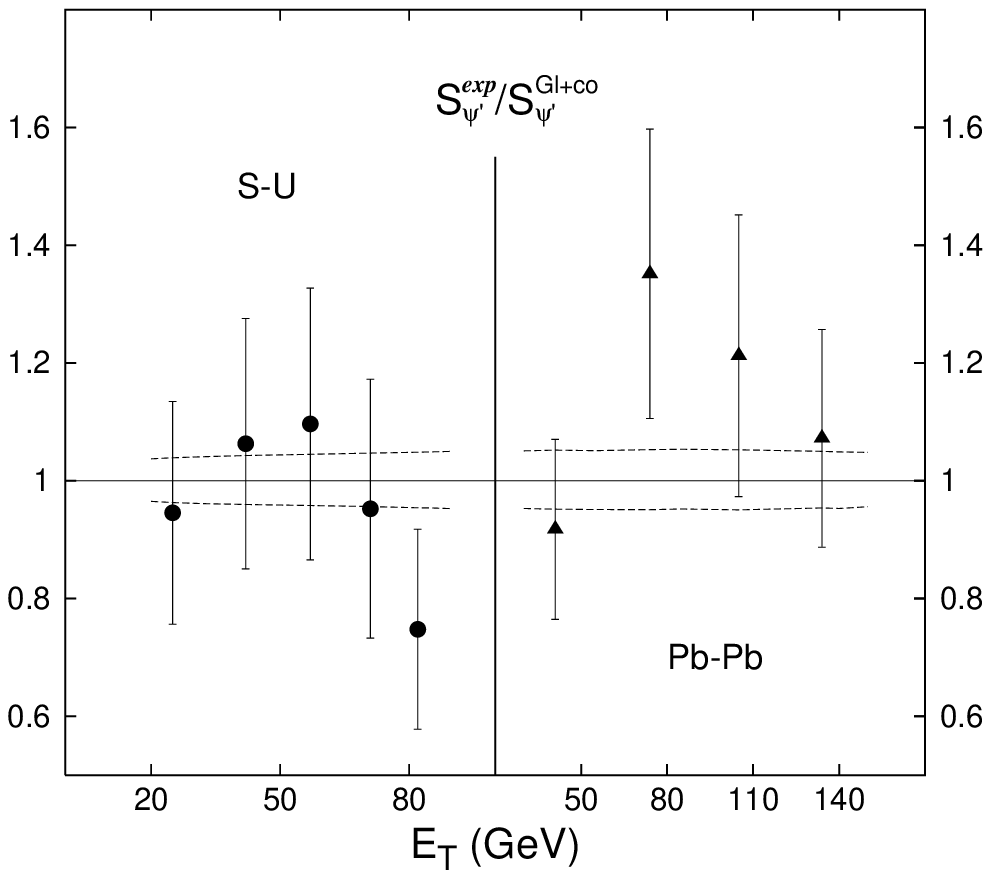,height=8cm}
\hspace{.6cm}
\begin{array}[b]{c}\left.\mathrm{b}\right)
\\ \\ \\ \\ \\ \end{array}
}\end{center}
\begin{center}\mbox{
\begin{array}[b]{c}\left.\mathrm{c}\right) 
\\ \\ \\ \\ \\ \end{array}
\hspace{.6cm}
\epsfig{file=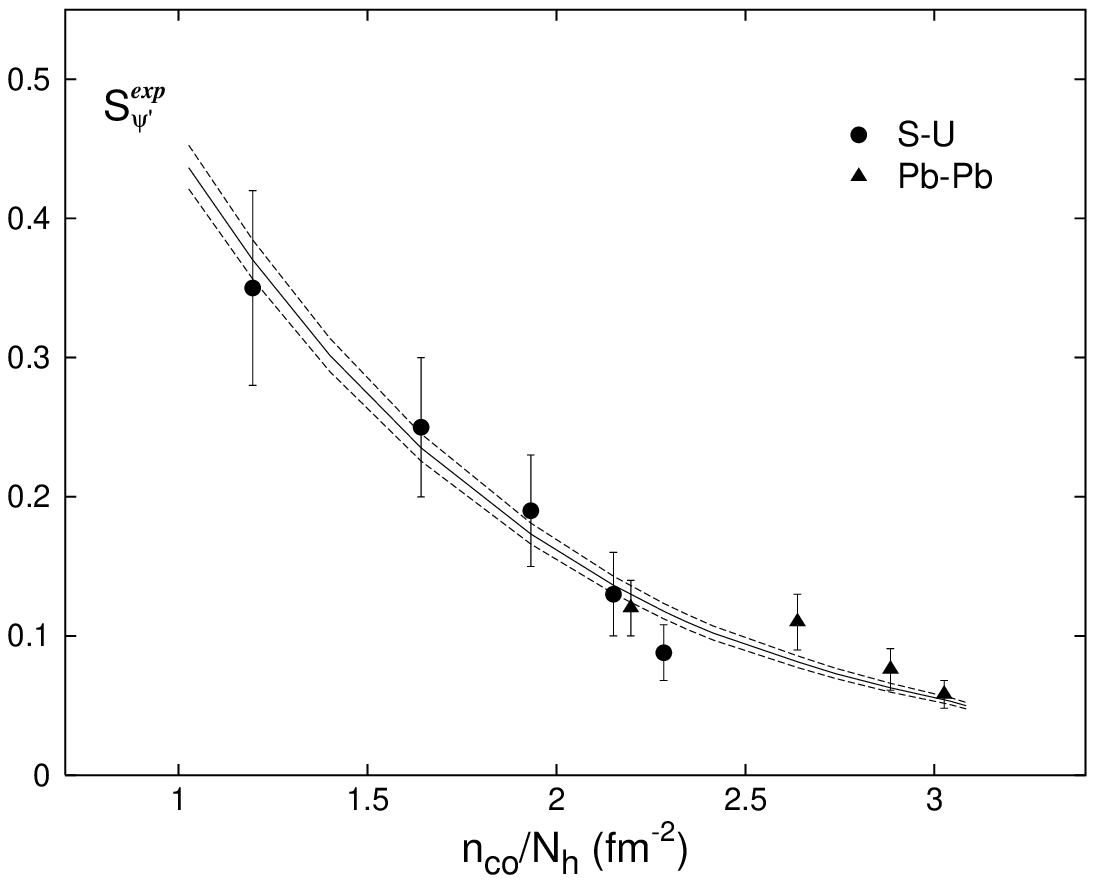,height=8cm}
\hspace{1.cm}
\begin{array}[b]{c}\left.\mathrm{ }\right.
\\ \\ \\ \\ \\ \end{array}
}\end{center}
Figure 8: The experimental $\psi^\prime$ survival probability 
compared to comover
absorption in $S-U$ (a) and $Pb-Pb$ (b) collisions; (c) shows
both data sets as function of $n_{co}$.
\newpage
\begin{center}\mbox{
\begin{array}[b]{c}\left.\mathrm{a}\right) 
\\ \\ \\ \\ \\ \end{array}
\hspace{.6cm}
\epsfig{file=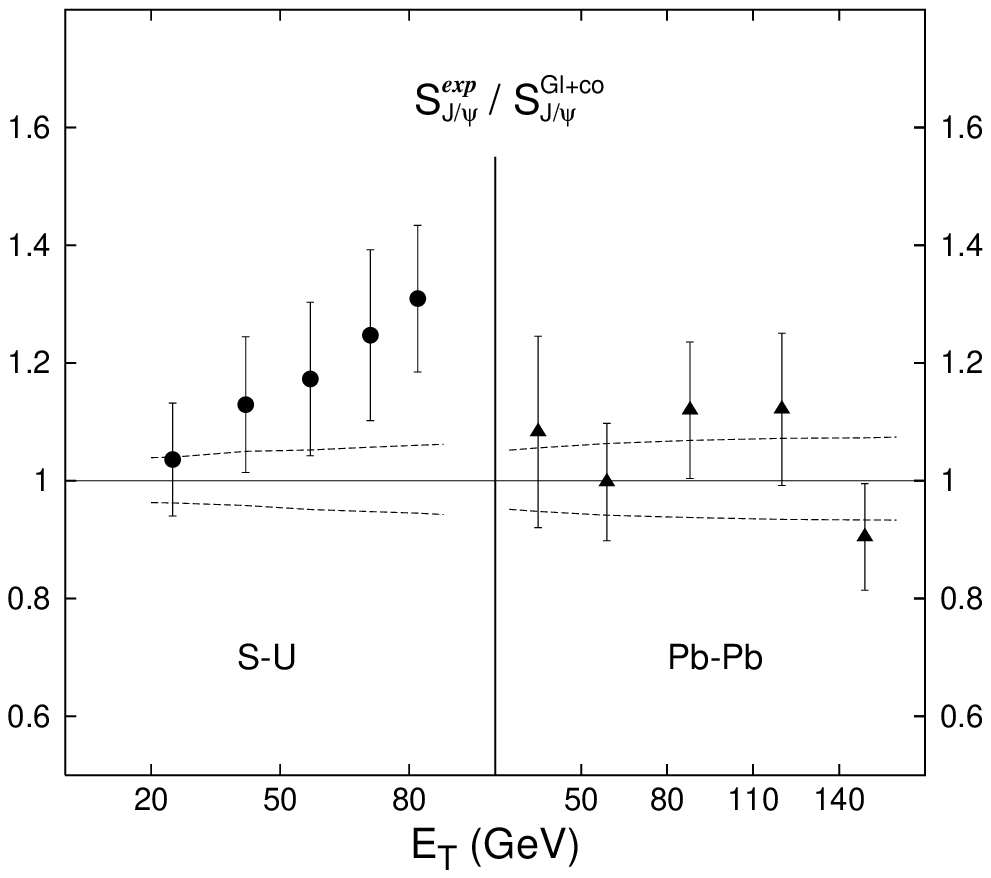,height=8cm}
\hspace{1.cm}
\begin{array}[b]{c}\left.\mathrm{ }\right.
\\ \\ \\ \\ \\ \end{array}
}\end{center}
\begin{center}\mbox{
\begin{array}[b]{c}\left.\mathrm{b}\right) 
\\ \\ \\ \\ \\ \end{array}
\hspace{.6cm}
\epsfig{file=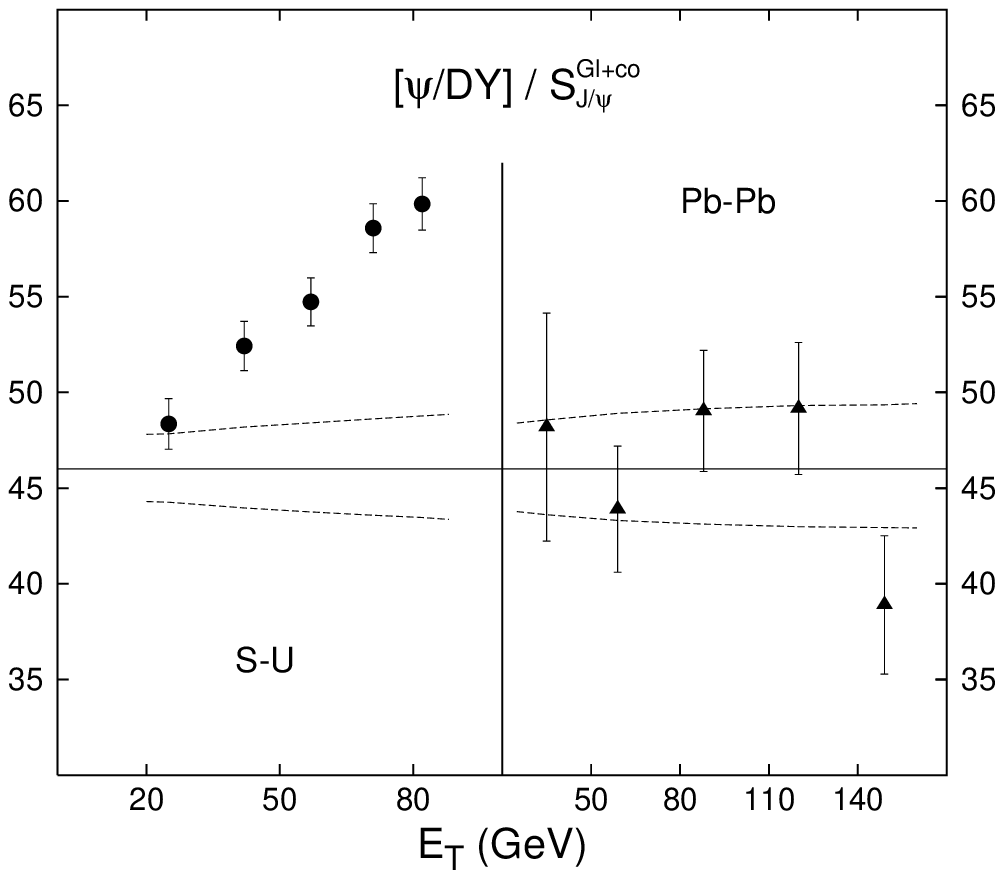,height=8cm}
\hspace{1.cm}
\begin{array}[b]{c}\left.\mathrm{ }\right.
\\ \\ \\ \\ \\ \end{array}
}\end{center}
Figure 9: The experimental $J/\psi$~survival probability (a) and the ratio 
of $J/\psi$~to Drell-Yan production (b) compared to comover absorption in 
$Pb-Pb$ and $S-U$ collisions.
\newpage
\begin{center}\mbox{
\epsfig{file=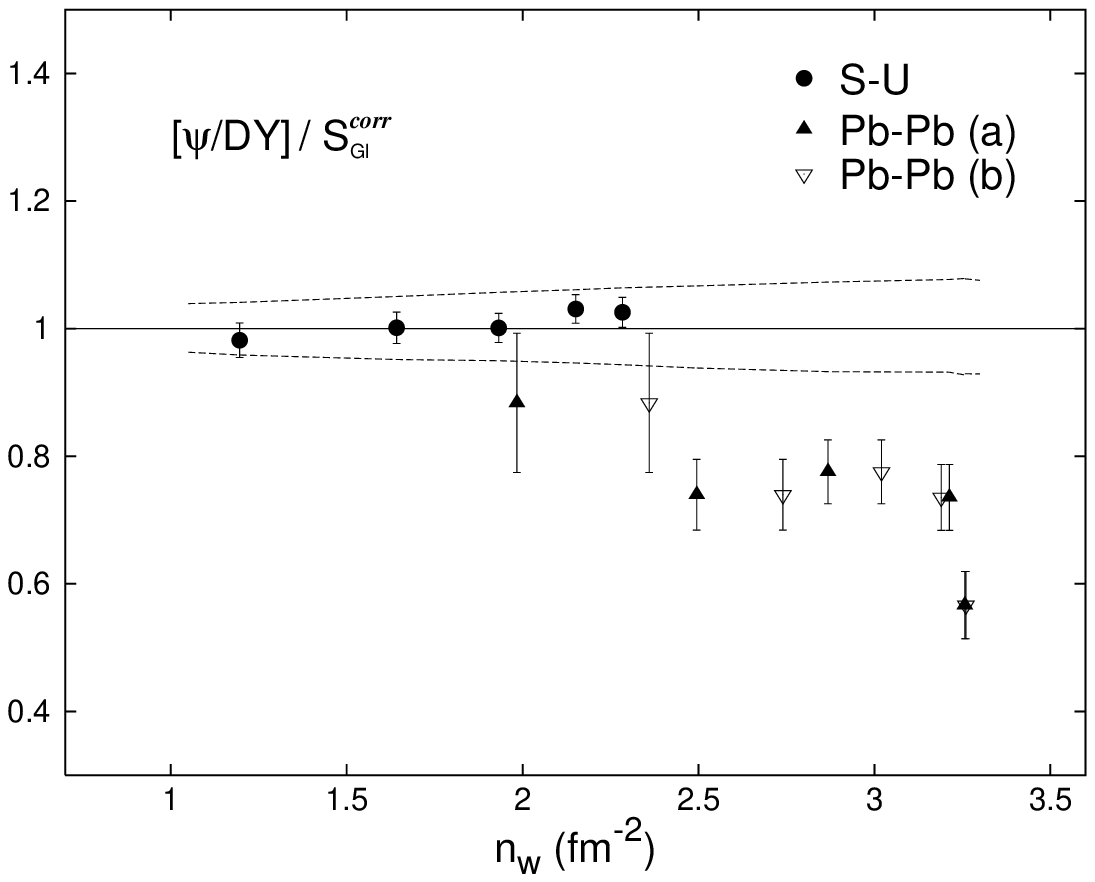,height=11cm}}\end{center}
Figure 10: The ratio of $J/\psi$~to Drell-Yan production in $S-U$ (circles) and
$Pb-Pb$ (triangles) collisions compared to ($\psi^\prime$-corrected)
Glauber theory results, as function of $n_w$; (a) filled triangles use
our Glauber impact parameter assignment, (b) open triangles that obtained 
from the experimental $E_T$-distribution \cite{Gonin}.
\newpage
\begin{center}\mbox{
\begin{array}[b]{c}\left.\mathrm{a}\right) 
\\ \\ \\ \\ \\ \end{array}
\hspace{.6cm}
\epsfig{file=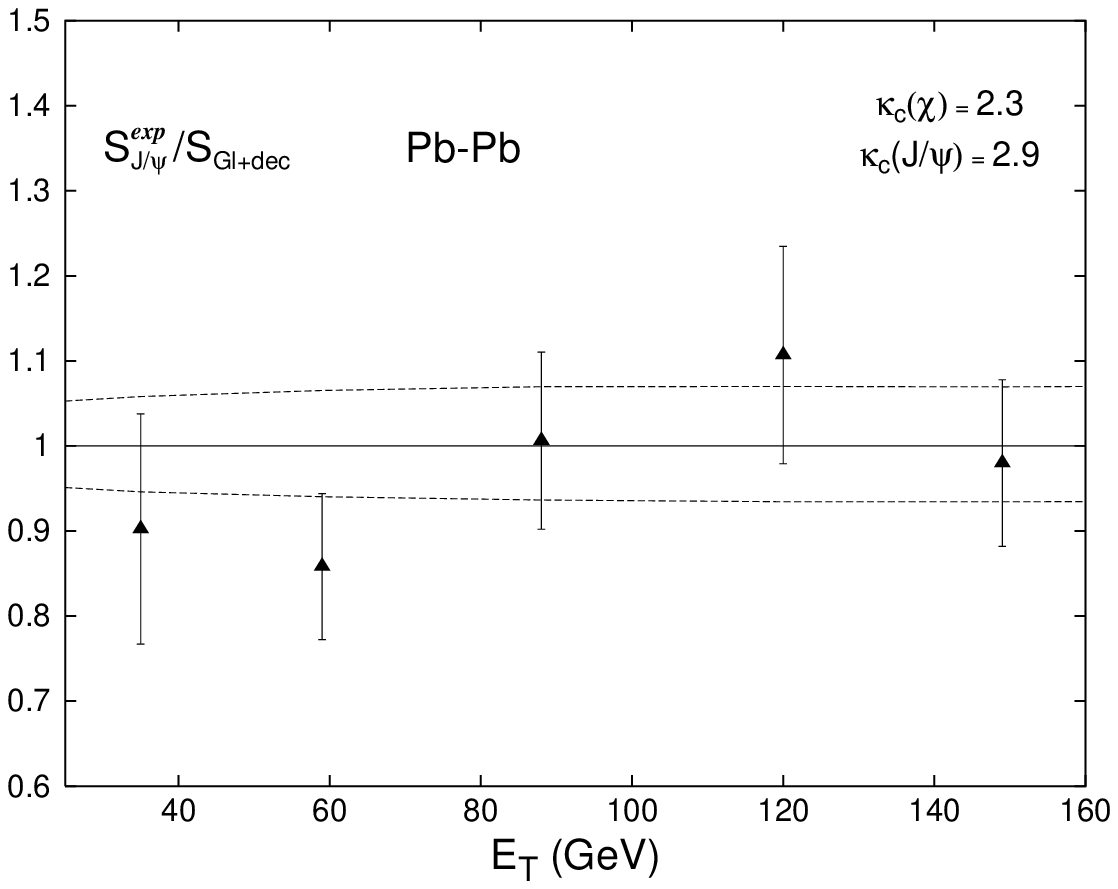,height=8cm}
\hspace{1.cm}
\begin{array}[b]{c}\left.\mathrm{ }\right.
\\ \\ \\ \\ \\ \end{array}
}\end{center}
\begin{center}\mbox{
\begin{array}[b]{c}\left.\mathrm{b}\right) 
\\ \\ \\ \\ \\ \end{array}
\hspace{.6cm}
\epsfig{file=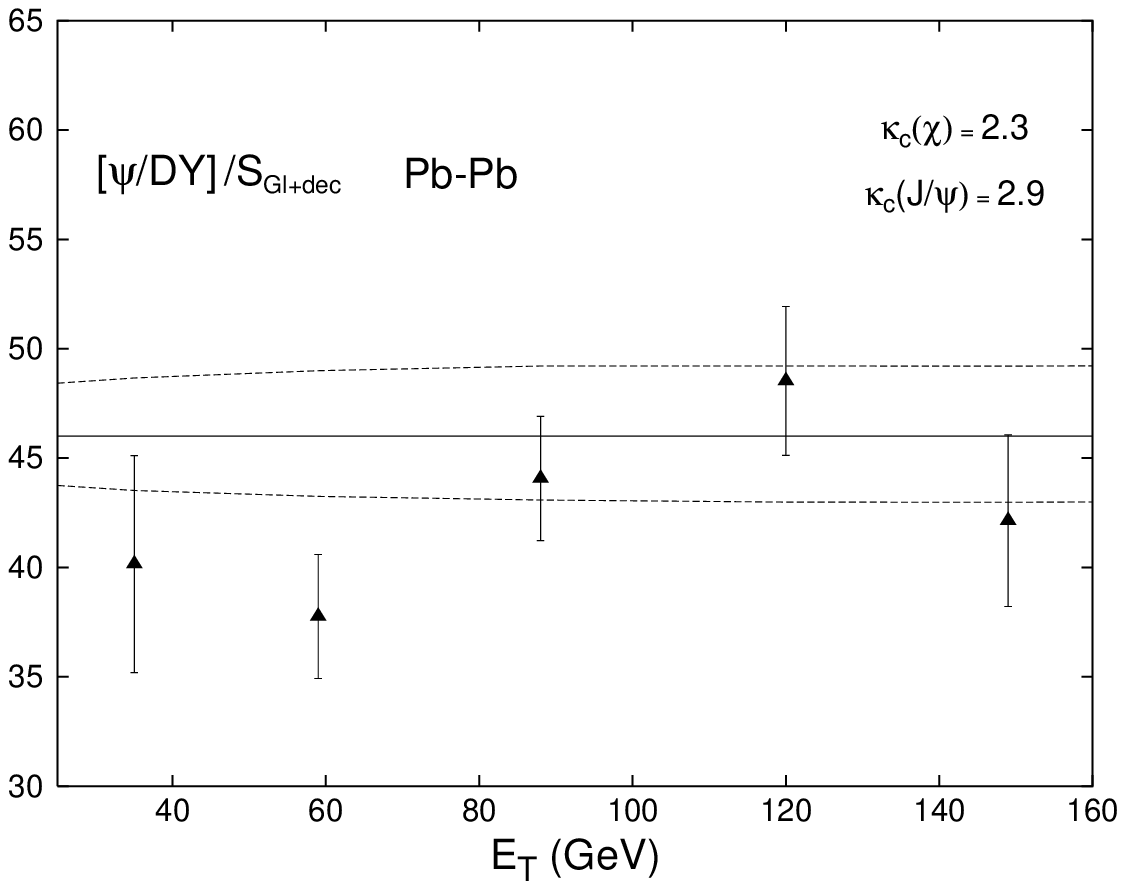,height=8cm}
\hspace{1.cm}
\begin{array}[b]{c}\left.\mathrm{ }\right.
\\ \\ \\ \\ \\ \end{array}
}\end{center}
Figure 11: (a) The experimental $J/\psi$~survival probability compared to
deconfinement suppression in $Pb-Pb$ collisions, with
$\kappa_c^{\chi}=2.3$ and $\kappa_c^{\psi}=2.9$; (b) same for
the ratio of $J/\psi$~to Drell-Yan production.
\newpage
\begin{center}\mbox{
\epsfig{file=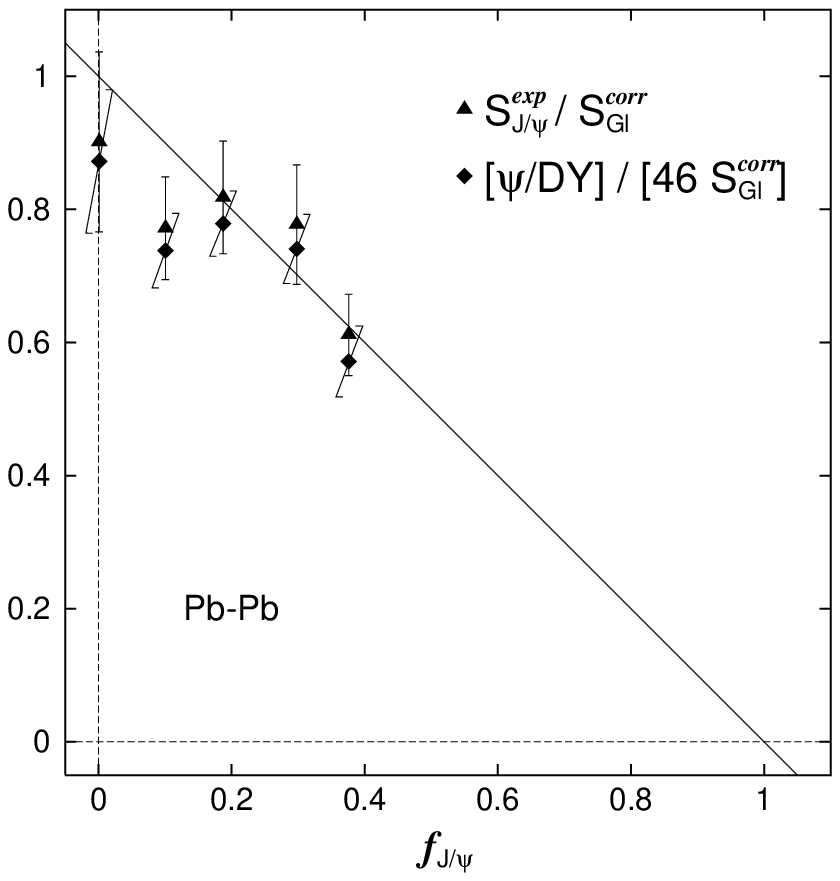,height=12cm}}\end{center}
Figure 12: Experimental $J/\psi$~suppression, compared to 
($\psi^\prime$-corrected)
Glauber theory results, as function of the
relative fraction $f_{J/\psi}$ of hot interaction volume.
\newpage

\end{document}